\documentclass[12pt]{article}
\usepackage{amssymb, amsmath, amsthm, amsfonts, latexsym, mathabx}
\usepackage{multirow} \usepackage{epsfig} \usepackage{psfrag} 
\usepackage{color}    \usepackage{xcolor} \usepackage{array} 
\usepackage[pdftex,bookmarks, colorlinks, citecolor =blue, breaklinks]{hyperref}
\usepackage{enumitem}\usepackage{bm}
\usepackage{longtable}

\newtheorem{prop}{Proposition}

\usepackage{graphicx}
\newcommand{\Zset}{\mathbb{Z}}  
  
\newcommand{\Rset}{\mathbb{R}}

\newcommand{\Sset}{\mathbb{S}}

\newcommand{\Fcal}{\mathcal{F}}
\newcommand{\Kcal}{\mathcal{K}}

\newcommand{\bfv}{{\bf v}}
\newcommand{\bfu}{{\bf u}}
\newcommand{\bfr}{{\bf r}}
\newcommand{\bfi}{{\bf i}}
\newcommand{\bfn}{{\bf n}}

\newcommand{\bfF}{{\bf F}}
\newcommand{\bfM}{{\bf M}}

\newcommand{\vp}{\varphi}
\newcommand{\vth}{\vartheta}
\newcommand{\al}{\alpha}

\newcommand{\hphi}{\hat{\phi}}

\begin{document}
\begin{center}
\noindent {\bf\Large Stable attitude dynamics of planar helio-stable and
drag-stable sails\footnote{This is based on the work presented in the AAS/AIAA
Astrodynamics Specialist Conference held in August 19-23, $2018$ in Snowbird,
Utah, U.S.A, published in the proceedings book as N. Miguel and C. Colombo,
{\sl Planar Orbit and Attitude Dynamics of an Earth-Orbiting Solar Sail under
$J_2$ and Atmospheric Drag Effects}, Advances in the Astronautical Sciences,
Vol. 167, 299-319, AAS 18-361.}}
\end{center}

\begin{center}
{\large Narc\'{\i}s Miguel and Camilla Colombo} 
\vspace*{1mm}

{\footnotesize
\noindent Dipartimento di Scienze e Tecnologie Aerospaziali\\
Politecnico di Milano\\
Via La Massa 34, 20156, Milano, Italia
}\\
 
\tt      {narcis.miguel@polimi.it}, 
\tt      {camilla.colombo@polimi.it} 
\end{center}

\begin{center}\today\end{center}

\begin{abstract}
In this paper the planar orbit and attitude dynamics of an uncontrolled
spacecraft is studied, taking on-board a deorbiting device. Solar and drag
sails with the same shape are considered and separately studied. In both cases,
these devices are assumed to have a simplified pyramidal shape that endows the
spacecraft with helio and drag stable properties. The translational dynamics is
assumed to be planar and hence the rotational dynamics occurs only around one
of the principal axes of the spacecraft. Stable or slowly-varying attitudes are
studied, subject to disturbances due to the Earth oblateness effect and gravity
gradient torques, and either solar radiation pressure or atmospheric drag
torque and acceleration. The results are analysed with respect to the aperture
of the sail and the center of mass - center of pressure offset.
\end{abstract}

\section*{Nomenclature}

{\renewcommand\arraystretch{1.0}
\noindent\begin{longtable}{@{}l @{\quad=\quad} l@{}}
$\alpha$ & aperture angle of the sail, deg or rad\\
$d$      & center of mass - center of pressure offset, m\\
$h$      & height of the panels, m\\
$w$      & width of the panels, m\\
$A_s$    & area of the panels, m$^2$\\
$m_s/2$  & mass of the panels, kg\\
$m_b$    & mass of the bus, kg\\
$\Fcal_b$ & body frame\\
$\xi,\eta,\zeta$ & coordinates of $\Fcal_b$\\
$\bfi_\xi, \bfi_\eta, \bfi_\zeta$ & unit vectors of the basis of $\Fcal_b$\\
$A,B,C$ & inertia moments of the whole spacecraft, kg m$^2$\\
$P$ & panels / parametrization of the panels in $\Fcal_b$\\
$\bfn$ & normal vectors to panels\\
$\Fcal_I$ & Earth centered inertial frame\\
$x, y, z$ & coordinates of $\Fcal_I$\\
$\bfi_x, \bfi_y, \bfi_z$ & unit vectors of the basis of $\Fcal_I$\\
$\phi,\Phi$ & Euler angle and angular velocity of the attitude, rad or deg, /s\\
$\lambda$ & Angle between Sun position and $\bfi_x$, rad or deg\\
$p_{\rm SR}$ & solar radiation pressure at 1 AU, N/m$^2$\\ 
$\delta$ & flight path angle, rad or deg\\
$\rho$ & atmospheric density, kg/m$^3$\\
$C_D$ & drag coefficient\\
$\bfu$ & unit vector\\
$\bfr, r$ & position vector and its magnitude, km\\
$\bfv, r$ & velocity vector and its magnitude, km/s\\
$\bfM, M$ & torque vector, component of torque vector, N m\\ 
$\bfF$ & force vector, N\\ 
$\sigma_{1,2,3}$ & cosines of the Earth-Sun vector in $\Fcal_b$\\
$\gamma_{1,2,3}$ & cosines of the Earth-spacecraft vector in $\Fcal_b$\\
$\nu_{1,2,3}$ & cosines of the relative velocity vector in $\Fcal_b$\\
$\Sigma$ & Surface of section where Poincar\'e iterates are computed\\
$a$ & semi-major axis of the spacecraft's orbit\\
$e$ & eccentricity of the spacecraft's orbit\\
$\Omega$ & Right ascension of the ascending node of the spacecraft's orbit\\
$\omega$ & argument of the perigee of the spacecraft's orbit, measured from ascending node\\
\end{longtable}}

\subsection*{Subscripts and upperscripts}
{\renewcommand\arraystretch{1.0}
\noindent\begin{longtable}{@{}l @{\quad=\quad} l@{}}
$\pm$ & that refers to panels $+$ or $-$\\ 
$S$   & that refers to Earth-Sun\\ 
${\rm sc}$  & that refers to spacecraft-Earth\\
${\rm rel}$ & that refers to the relative velocity with respect to the atmosphere\\
$\xi,\eta,\zeta$ & in the direction of, in $\Fcal_b$\\
$x, y, z$ & in the direction of, in $\Fcal_I$\\
\end{longtable}}

%---------------------------------------------------------------------- 
% Section: Introduction
%----------------------------------------------------------------------
%------------------------------------------------------------
% Introduction
%------------------------------------------------------------
\section{Introduction}

Solar sails are a low-thrust propulsion that relies on the Solar Radiation
Pressure (SRP). They have attracted much attention in the literature, since a
spacecraft with a solar sail generated acceleration in a slow but continuous
way allowing to reduce the cost of missions. This technology has been
successfully demonstrated in various missions, see for instance JAXA's
IKAROS~\cite{Tsu11}, the Planetary Society's LightSail projects and NASA's
NanoSail-D project~\cite{Joh11}. The latter demonstrated the feasibility of the
deployment of a sail and its usage to deorbit a spacecraft exploiting the
effect of atmospheric drag.

There is a vast literature on how to use the enhancements of the effects of
SRP and drag for mission design. A common feature among these works is to
assume that, along the trajectories, the attitude of the sail is fixed; hence
the feasibility of these works rely on attitude control.

\begin{figure}[h!]
\begin{center}
\includegraphics[width = 0.48\textwidth]{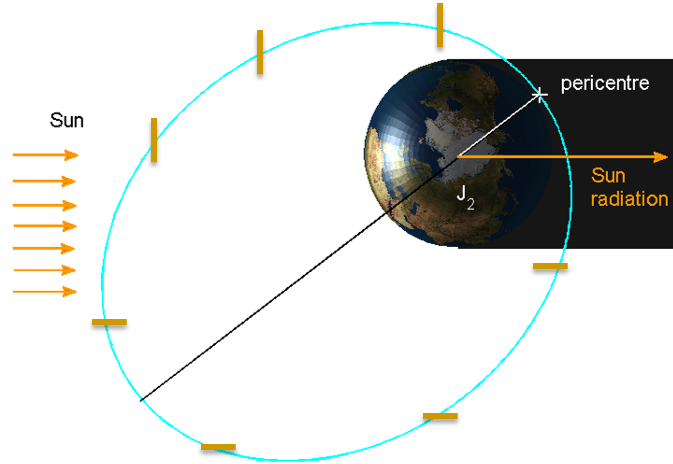}
\includegraphics[width = 0.48\textwidth]{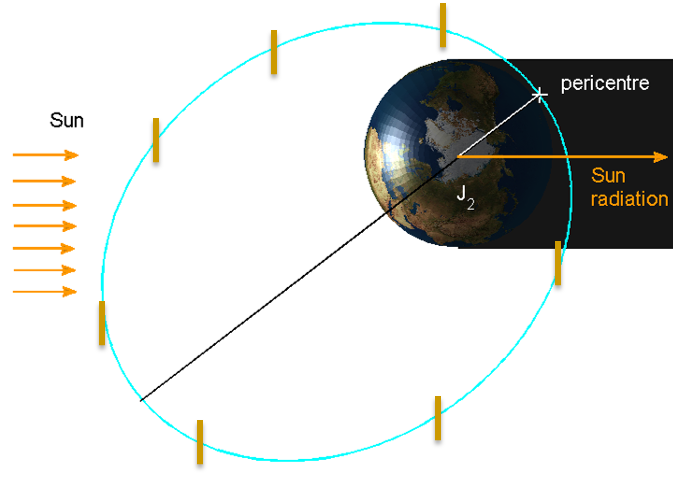}
\end{center}
\vspace{-0.5cm}
\caption{Sail orientation in the active (left) and the passive (right)
deorbiting strategies~\cite{CBF16}.}
\label{fig:actvspas}
\end{figure}

In this work and we build on studies whose objectives are end-of-life disposals
employing sails as passive deorbiting devices. For deorbiting from an altitude
where atmospheric drag is the dominant effect, the sail can be either
controlled, to keep always its maximum cross area perpendicular to the incoming
air flow, or uncontrolled and therefore tumbling. In this case the cross area
exposed to aerodynamic drag will be varying in time~\cite{CRVBBK17}. For orbit
altitude above 800 km, the effect of SRP can be exploited to achieve
deorbiting.  Deorbiting strategies making use of SRP can be splitted in two
main attitude control strategies, active and passive, as defined
in~\cite{CBF16}. Active strategies allow deorbiting ``inwards" on a spiraling
path by decreasing the semi-major axis of the orbit. This is achieved by
maximizing the SRP effect when approaching the Sun and minimizing it when
moving away from the Sun~\cite{BT06}, see the left panel in
Figure~\ref{fig:actvspas}. On the other hand, the passive approach requires a
fixed attitude of the spacecraft with respect to the Sun, and it consists of
the counter-intuitive idea of deorbiting ``outwards" by increasing the
eccentricity of the orbit, since this implies the decrease of the
perigee~\cite{LCI12, LCI13}, see the right panel in Figure~\ref{fig:actvspas}. 

Here the following natural question arises: can one find a sail with
auto-stabilizing properties, so that the already cited strategies can apply
minimizing the need for attitude control? The answer is affirmative from the
point of view of SRP, and it is achieved by means of a Quasi-Rhombic Pyramid
(QRP) shape, as suggested in~\cite{CHMR13}. The structure is formed by 4
reflective panels resembling the shape of the pyramid. If oriented towards the
sunlight, such a structure is expected to compensate, on average, the
components of the acceleration in any other direction. Namely, in~\cite{FCH16}
the authors provide a first-order (and hence local) argument for the stability
of the sun-pointing attitude, and they later study the possible stability
enhancements of assuming a moderate spin around this direction in~\cite{FHC17}.

Despite the authors of~\cite{CHMR13, FCH16, FHC17} obtain satisfactory results
by considering such structure, there is, to the author's knowledge, a lack of
understanding on the stability from a more global point of view. That is, if
there are also stable attitude dynamics close to the sun-pointing orientation,
and, in affirmative case, if one can measure and describe the set of stable
motion. This paper is a first step in this direction.\\

The goal of this paper is to give evidence of the possibilities of QRP as
feasible auto-stabilized deorbiting devices both in SRP dominated regions and
in atmospheric drag dominated regions, that are studied separately. The motion
is assumed to be planar -the obliquity of the ecliptic is set to zero and the
rotations occur around an axis perpendicular to the orbital plane- and the QRP
is simplified so that out-of-plane motion is avoided. Also, the effect of
eclipses is neglected. The attitude stability study takes into account two main
parameters: aperture angle of the sail structure, $\alpha$ and the center of
Mass - center of Pressure Offset (MPO), that is a signed real variable $d$.\\

The paper is structured as follows. First of all \S~\ref{sec:geom} is devoted
to the study of the geometry of the spacecraft under consideration and to
provide their inertia moments taking into account the parameters $\alpha$ and
$d$. This is used in \S~\ref{sec:model} to provide explicit expressions for the
SRP, atmospheric drag and gravity gradient torques. This allows to set the
equations of motion to be studied. 

The next two sections are devoted to investigate attitude stability of the
family of spacecraft under consideration in SRP dominated regions and in
atmospheric drag dominated regions.
\begin{enumerate}
	\item On the one hand, the hypotheses considered allow to start
\S~\ref{sec:SRP} by approximating the dynamics as a one and a half degrees of
freedom Hamiltonian system that allows to study separately SRP and gravity
gradient torques as main effect and perturbation, respectively. This allows, in
particular, to establish necessary physical relations between $\al$ and $d$ for
helio-stability. The model is validated comparing the results with the
integration of the full coupled orbit and attitude model in two test cases.
	\item The analogous problem taking into account drag instead of SRP is
considered in \S~\ref{sec:DRAG}. The differences and common features between
both scenarios are first compared and a numerical study of the performance of
the spacecraft under consideration as orbiting devices is performed. 
\end{enumerate}

This contribution ends in \S~\ref{sec:conclusions} with a summary of the
obtained results, conclusions and future lines of research that emerge from
this paper.

%---------------------------------------------------------------------- 
% Section: Geometry of the sail
%----------------------------------------------------------------------
%-----------------------------------------------------------
% Geometria de la vela 
%-----------------------------------------------------------
\section{Geometry of the sail}\label{sec:geom}

\begin{figure}[h!]
\begin{center}
\includegraphics[width = 0.45\textwidth]{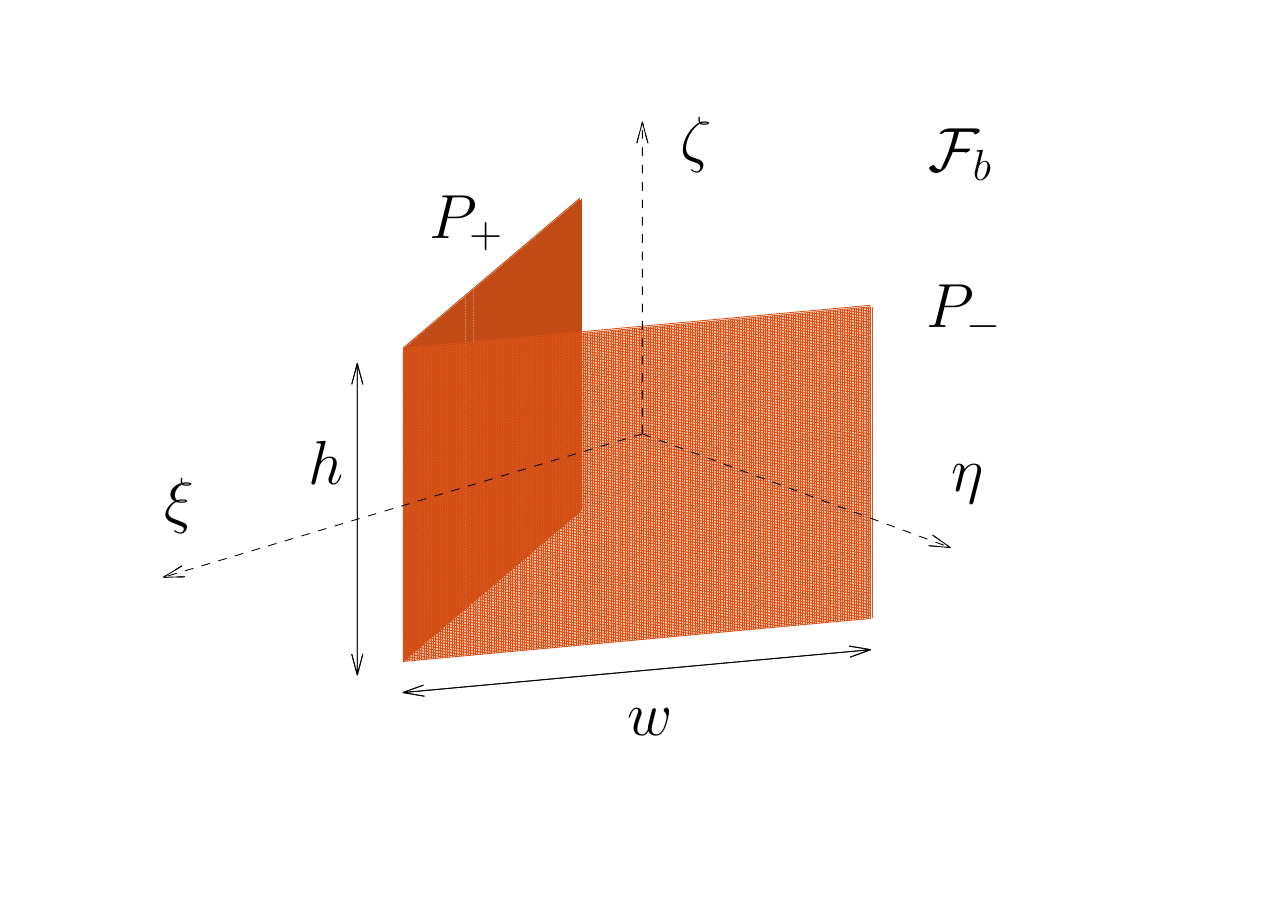}
\includegraphics[width = 0.45\textwidth]{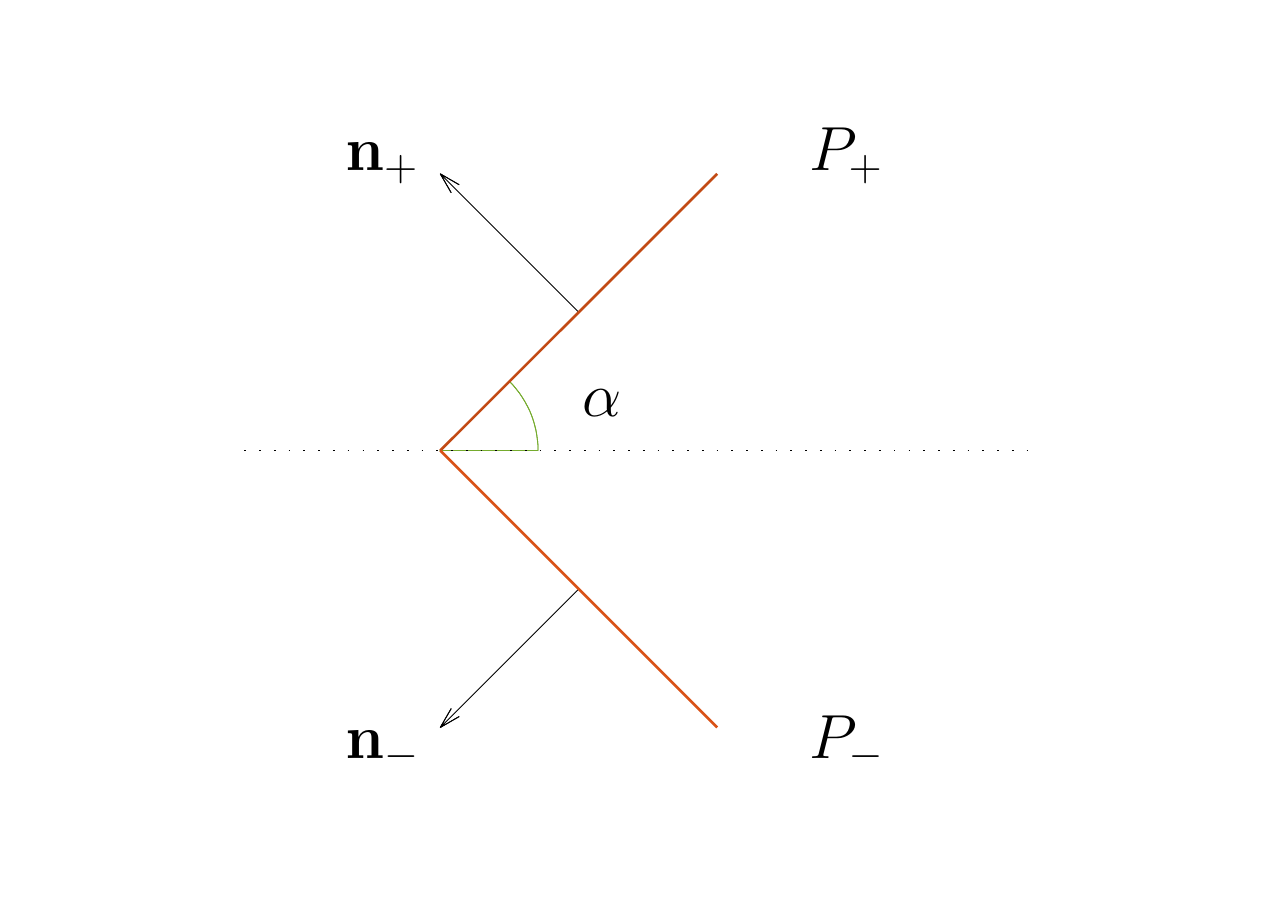}
\end{center}
\caption{Sketch of the sail structure. Left: 3D view. Right: top view.}
\label{fig:shapesail}
\end{figure}

To avoid out-of-plane motion one is lead to consider a simplification of a QRP
that consists of two panels of equal size; say of height $h$, width $w$, and
area $A_s=hw$. Assume that the weight of each panel is $m_s/2$, so the mass of
the whole sail structure is $m_s$. In the left panel of
Fig.~\ref{fig:shapesail} a sketch of the sail structure is depicted. The right
panel is a top view of the left sketch.

%- Figura

The parametrisation of the sail panels is written in a reference frame
$\Fcal_b$ attached to the spacecraft whose coordinates are $\xi,\eta$ and
$\zeta$. Call $\bfi_{\xi, \eta,\zeta}$ the vectors of the basis. In the
cylindrical coordinates $(r, \alpha, \zeta)$, $r = \xi^2+\eta^2$ and
$\tan\alpha = \eta / \xi$, the panels of the sail are considered to be
parametrized as
\begin{eqnarray}\label{eq:paramsail}
P_+\cup P_-,&&
P_\pm = \left\{
\left({\tt aux} - r\cos\alpha, \pm r\sin\alpha, \zeta\right)^\top:
      \;r\in[0, w],\;\zeta\in[-h/2, h/2]
\right\},
\end{eqnarray}
where ${\tt aux}$ is a free parameter that will be chosen so that the center of
mass of the whole spacecraft is at the origin of $\Fcal_b$. The panels are
attached to each other along an $h$-long side, that lies on a line parallel to
the $\zeta$ axis, and they form an angle $\alpha$ with respect to the plane $\eta =
0$. Assuming uniform density of the panels, the centre of mass of the structure
is at
$$
\bfr_s = \left({\tt aux} - \frac{1}{2}w\cos\alpha, 0, 0 \right)^\top.
$$

\begin{figure}[h!]
\begin{center}
\includegraphics[width = 0.32\textwidth]{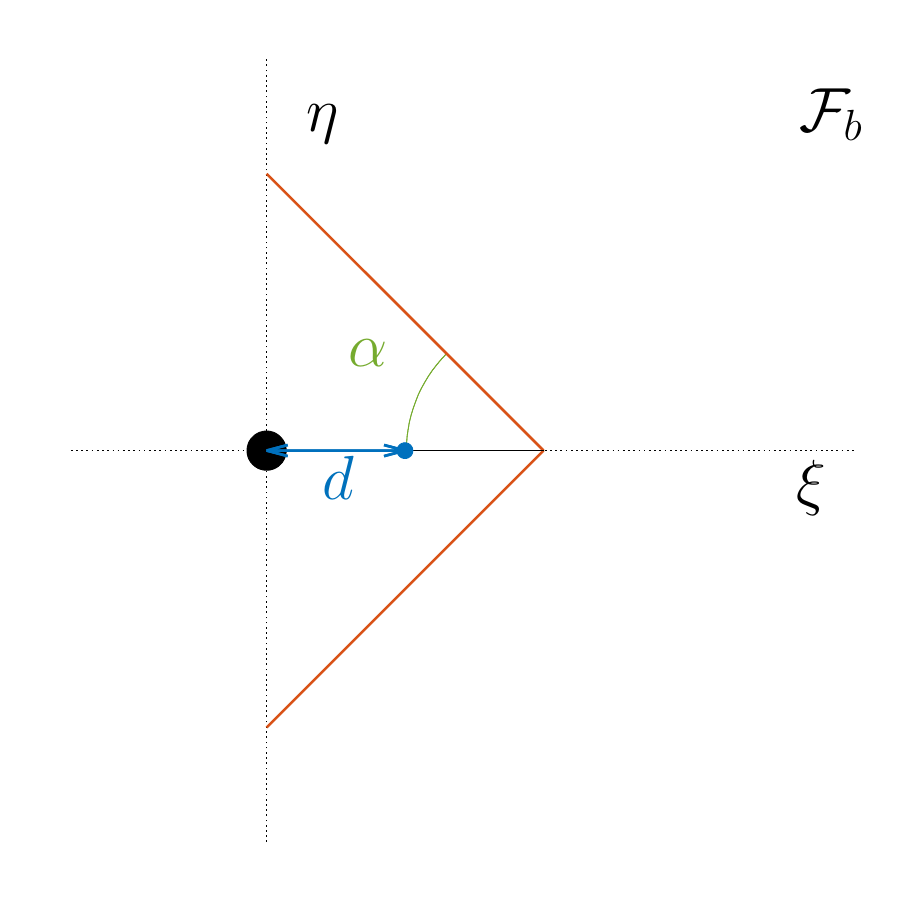}
\includegraphics[width = 0.32\textwidth]{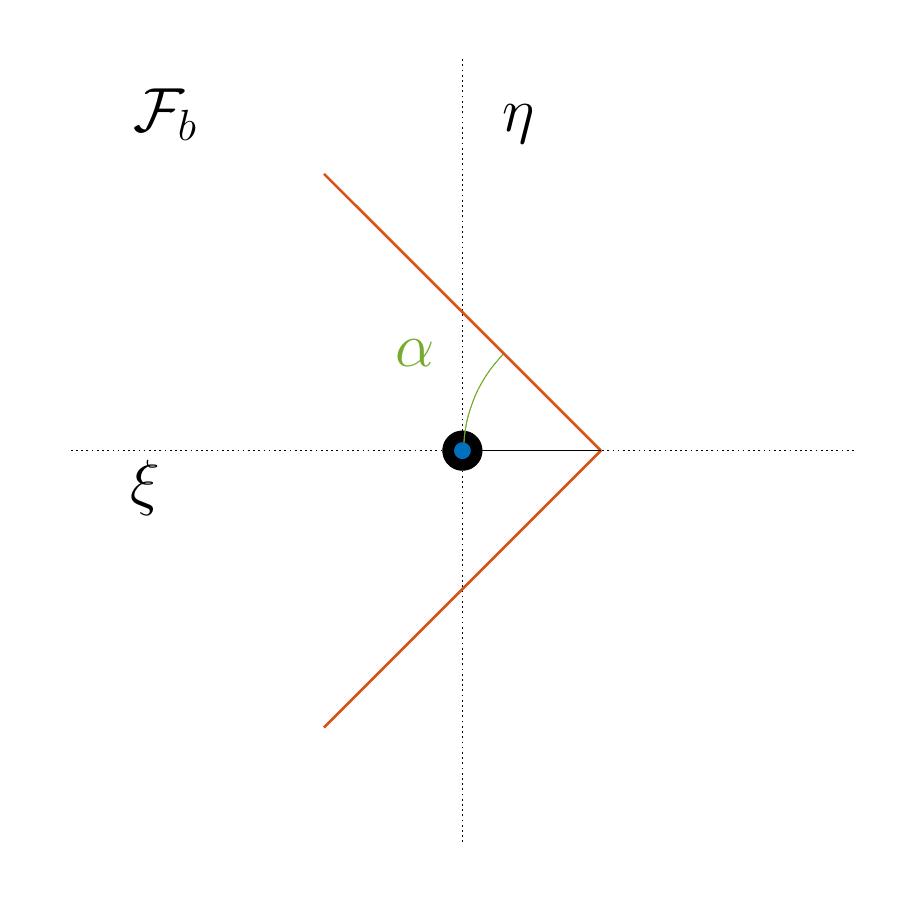}
\includegraphics[width = 0.32\textwidth]{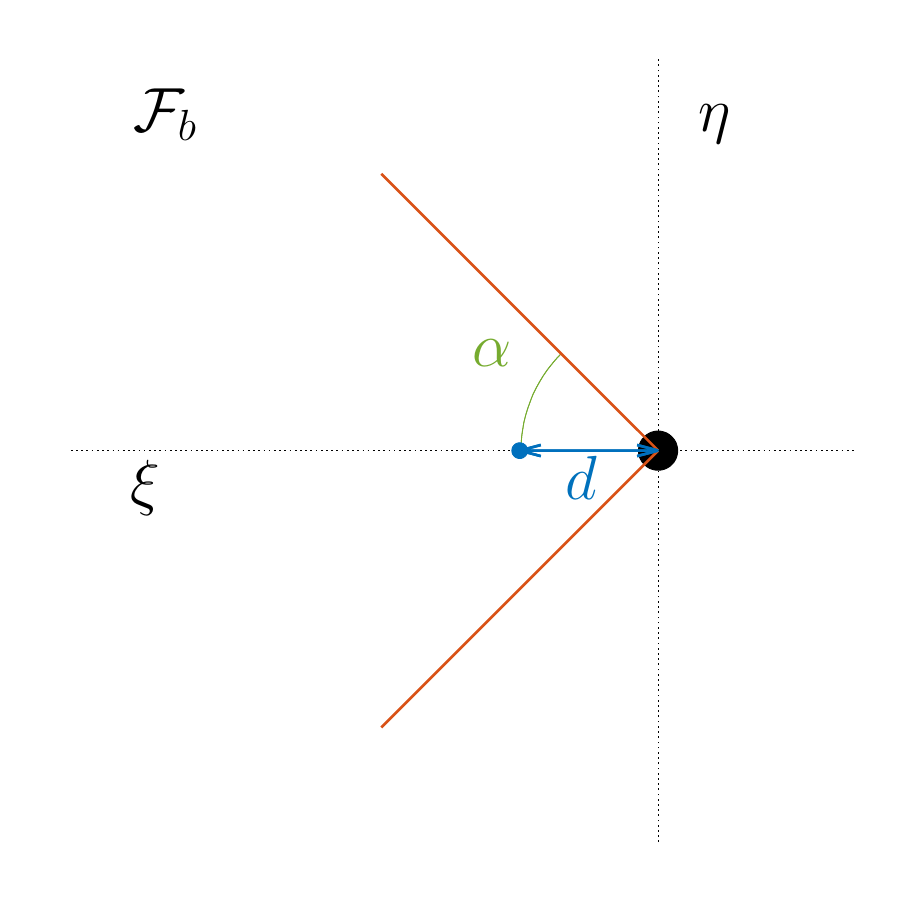}
\end{center}
\caption{Sketch of the top view of the spacecraft in $\Fcal_b$, where the bus
is depicted as a solid circle. Left: $d < 0$. Center: $d = 0$. Right: $d > 0$.}
\label{fig:shapesc}
\end{figure}

Note that chosen this way, the principal axes of inertia of the sail are
parallel to those of $\Fcal_b$, and this remains true if the center of mass of
the bus of the satellite lies on the $\xi$ axis. So, assume that the latter is
located at the point $\bfr_s + (d, 0, 0)^\top$, $d\in\Rset$. The parameter $d$
accounts for the MPO, and its sign informs about the relative position of the
payload with respect to the sail. The centre of mass of the whole spacecraft is
the origin if
\begin{eqnarray}\label{eq:cdmeq0}
{\tt aux} & = & \frac{1}{2}w\cos\alpha - d\frac{m_b}{m_b + m_s},
\end{eqnarray}
where $m_b$ is the mass of the bus of the spacecraft. Sketches
of top views of spacecraft in $\Fcal_b$ can be seen in Fig.~\ref{fig:shapesc},
where the bus is depicted as a solid black dot. The left, center and right
panels are sketches of spacecraft with $d<0$, $d=0$ and $d>0$, respectively.
Note that, in particular, the bus is placed at the tip of the sail (where both
panels are in contact) if we choose $d$ such that ${\tt aux} = 0$ in
Eq.~\ref{eq:cdmeq0}, that is,
\begin{eqnarray}\label{eq:busattip}
d & = & \frac{1}{2}w\cos(\al)\frac{m_b+m_s}{m_b}.
\end{eqnarray} 
%- Figura

Assume that the principal axes of the payload are also parallel to the axes of
$\mathcal{F}_b$, and denote $I_{\xi,b}, I_{\eta,b}$ and $I_{\zeta,b}$ its
moments of inertia if its centre of mass is at the origin. Using the parallel
axes theorem see, e.g.~\cite{Gol80}, the moments of inertia along the $\xi$,
$\eta$ and $\zeta$ axes of the whole spacecraft are, respectively,
\begin{subequations}
\begin{eqnarray}
&&A = I_{\xi,b} + \frac{h^2 m_s}{6},
\quad 
B = I_{\eta,b} + \frac{h^2 m_s}{6} + D(\alpha,d),
\quad
C = I_{\zeta,b} + D(\alpha,d), \label{eq:momine}\\
&&D(\alpha,d)  =  
\frac{1}{6}m_sw^2\cos^2\alpha +
\frac{d^2m_b^2(m_b+2m_s)}{(m_b+m_s)^2}\label{eq:Dald}
\end{eqnarray}
\end{subequations}

Finally, denote $\bfn_\pm = (\sin\alpha,\pm\cos\alpha,0)$ the normal vector of
the panels $P_\pm$, see the right sketch in Fig.~\ref{fig:shapesail}.

%---------------------------------------------------------------------- 
% Section: Model
%----------------------------------------------------------------------
%-----------------------------------------------------------
% Model
%-----------------------------------------------------------
\section{Model of planar orbit and attitude dynamics}\label{sec:model}

The planar orbit and attitude dynamics considered here is a coupled system of
differential equations in $(\Sset^1\times\Rset)\times\Rset^4$, where
$\Sset^1:=\Rset/(2\pi\Zset)$: orientation and angular velocity of the
rotational dynamics in $\Fcal_b$ and position and velocity of the spacecraft in
an Earth centered inertial frame $\Fcal_I$.  Denote the coordinates of
$\Fcal_I$ $x,y$ and $z$, and the vectors of the orthonormal basis
$\bfi_{x,y,z}$. The vector $\bfi_x$ points towards an arbitrarily chosen
direction on the ecliptic (e.g.  J2000), and since the dynamics in this paper
is restricted to the plane, the vectors $\bfi_z$ and $\bfi_\zeta$ are parallel.
The triad is completed by choosing $\bfi_y = \bfi_z\times \bfi_x$.

This section deals with vectors in the two frames $\Fcal_I$ and $\Fcal_b$. To
avoid confusion and unless a formula that applies in both frames is given, the
subscript $I$ and $b$ refer to vectors in $\Fcal_I$ and $\Fcal_b$,
respectively.

In the case planar motion, the rotation dynamics of the spacecraft is fully
explained using a single Euler angle, $\vp\in[0, 2\pi)$. Assume the change of
coordinates from $\Fcal_I$ to $\Fcal_b$ is done through $R_3(-\vp)$, where 
$$
R_3(\psi) = 
\left(
\begin{array}{rrc}
 \cos\psi&-\sin\psi&0\\
 \sin\psi& \cos\psi&0\\
0&0&1
\end{array}
\right).
$$
The Euler equations in this situation reduce to
\begin{eqnarray}\label{eq:EulerSimp}
C\ddot{\vp} & = &M_\zeta
\qquad\mbox{ or }\qquad
\left\{
\begin{array}{rcl}
\dot{\vp}  & = & \Phi\\
\dot{\Phi} & = & M_\zeta/C
\end{array}
\right.,
\end{eqnarray}
where $\Phi$ is the rotational angular velocity and $C$ is the third inertia
moment, recall Eq.~\ref{eq:momine}, and $M_\zeta$ refers to the sum of the
components along the $\zeta$ direction of the torques under consideration, that
will be derived in \S~\ref{subsec:perturbations}. The state vector of the
complete problem is of the form
$$
\left[\vp,\; \Phi,\; x,\; y,\; v_x,\; v_y\right]^\top .
$$

%-----------------------------------------------------------
% Orbit and attitude perturbations 
%-----------------------------------------------------------
\subsection{Orbit and attitude perturbations}\label{subsec:perturbations}

Assume that the apparent motion of the Sun around the Earth is circular with
constant angular velocity $n_\odot$, and let $\lambda$ denote the angle of
position of the Sun on the orbital plane with respect to $\bfi_x$. In the
frames $\Fcal_{I}$ and $\Fcal_{b}$ the Earth-Sun vector reads
\begin{eqnarray}\label{eq:earthsun}
\bfu_{S,I} & = & (\cos\lambda,\sin\lambda,0)^\top_I,\quad \bfr_{S,I} = r_S \bfu_{S,I};
\quad
\bfu_{S,b}  =  R_3(-\vp)\bfu_{S,I}, \quad \bfr_{S,b} = r_S \bfu_{S,b}.
\end{eqnarray}

\noindent On the other hand, the Earth-spacecraft vector in the $\Fcal_{I}$ and
$\Fcal_{b}$ frames read
\begin{eqnarray}\label{eq:earthsc}
\bfu_{{\rm sc},I} & = & (\cos\vth,\sin\vth,0)^\top_I,\quad \bfr_{{\rm sc},I} = r_{\rm sc} \bfu_{{\rm sc},I};
\quad
\bfu_{{\rm sc},b}  =  R_3(-\vp)\bfu_{{\rm sc},I}, \quad \bfr_{{\rm sc},b} = r_{\rm sc} \bfu_{{\rm sc},b},
\end{eqnarray}
where $r_{\rm sc}=\sqrt{x^2+y^2}$ and $\vth = \Omega + \omega + \theta$, being
$\Omega$ the Right Ascension of the Ascending Node (RAAN)\footnote{Here the
RAAN is taken into account as the $J_2$ perturbation causes the precession of
the line of nodes.}, $\omega$ is the argument of the perigee and $\theta$ is
the true anomaly of the osculating orbit.

\begin{figure}[h!]
\begin{center}
\includegraphics[width = 0.48\textwidth]{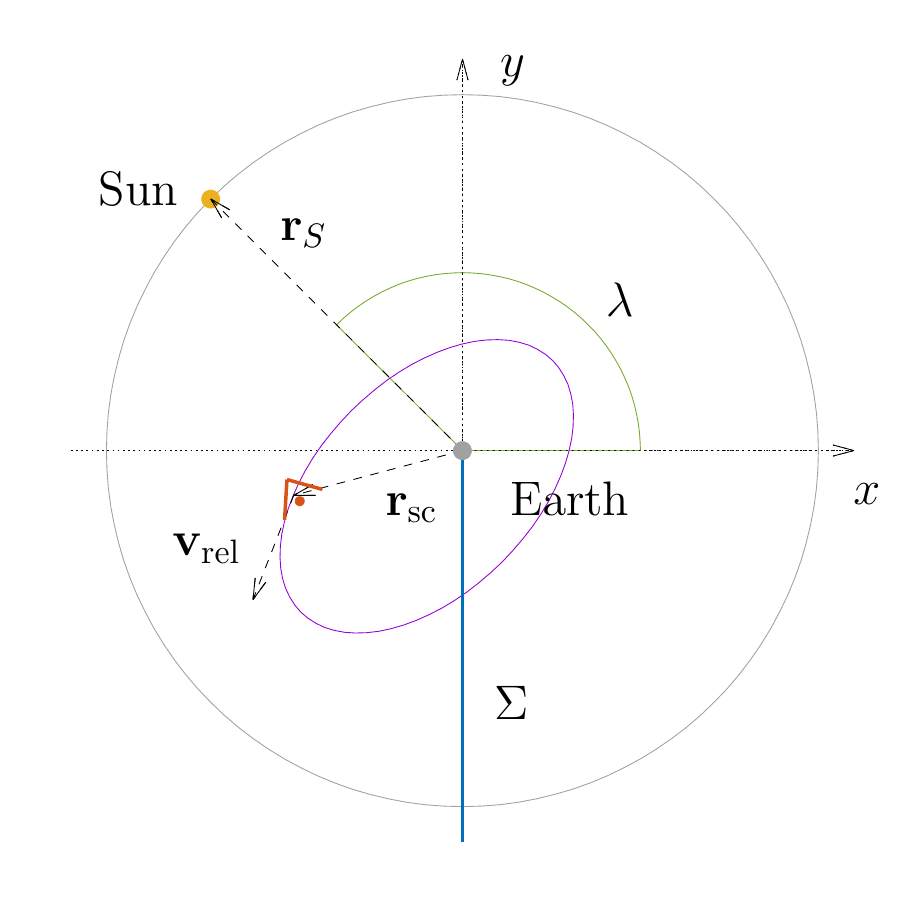}
\end{center}
\caption{Sketch of the main elements that play a role in the dynamics of the
studied family of spacecraft.}
\label{fig:sketchSigma}
\end{figure}

Finally the relative velocity of the spacecraft with respect to the atmosphere
reads, in the $\Fcal_{I}$ and $\Fcal_{b}$ frames
\begin{eqnarray}\label{eq:vrel}
\bfu_{{\rm rel},I} = (\cos\delta, \sin\delta ,0)^\top_I,\quad \bfv_{{\rm rel},I} = v_{\rm rel} \bfu_{{\rm rel},I};
\quad
\bfu_{{\rm rel},b}  =  R_3(-\vp)\bfu_{{\rm rel},I}, \quad \bfv_{{\rm rel},b} = v_{\rm rel} \bfu_{{\rm rel},b},
\end{eqnarray} 
where $\delta:=\arctan(v_y/v_x)$ and $v_{\rm rel} =
\sqrt{v_x^2 + v_y^2}$. In Fig.~\ref{fig:sketchSigma} a sketch of the considered
motion and the vectors $\bfr_S, \bfr_{\rm sc}$ and $\bfv_{\rm rel}$ is shown.

% - Figura

For convenience, let us denote 
\begin{subequations}\label{eq:vectors}
\begin{eqnarray}
\bfu_{S,b}         & := &\sigma_1 \bfi_{\xi} + \sigma_2 \bfi_{\eta} + \sigma_3 \bfi_\zeta,\label{eq:uS}\\
\bfu_{{\rm sc},b}  & := &\gamma_1 \bfi_{\xi} + \gamma_2 \bfi_{\eta} + \gamma_3 \bfi_\zeta,\label{eq:uE}\\
\bfu_{{\rm rel},b} & := &\nu_1    \bfi_{\xi} + \nu_2    \bfi_{\eta} + \nu_3    \bfi_\zeta,\label{eq:urel}
\end{eqnarray}
\end{subequations}
where
$$
\sigma_1^2 + \sigma_2^2 + \sigma_3^2 \; = \; 
\gamma_1^2 + \gamma_2^2 + \gamma_3^2 \; = \; 
\nu_1^2    + \nu_2^2    + \nu_3^2    \; = \; 1
$$ 
are direction cosines.\\ 

The shape of the sail structure in Fig.~\ref{fig:shapesail} makes the torques
due to SRP and drag acceleration have a different representation depending on
how the sail is oriented with respect to the sunlight or relative velocity
vectors, respectively. Since the scope of this paper are stable motions close
to either the sun-pointing or velocity-pointing directions, we are only lead to
consider the following cases: 
\begin{enumerate}
	\item For SRP, denote $\phi = \vp - \lambda$. If $\phi \in (-\alpha,
\pi-\alpha)$, the panel $P_-$ produces acceleration and torque; and if $\phi
\in (-\pi+\alpha, \alpha)$, the panel $P_+$ does; in particular, if
$|\phi| < \alpha$, both panels face the sunlight.  
	\item For atmospheric drag, let $\hphi = \vp - \delta$. Similarly, if
$\hphi \in (-\alpha, \pi-\alpha)$, the panel $P_-$ produces acceleration and
torque, if $\hphi \in (-\pi+\alpha, \alpha)$, the panel $P_+$ does, and for
$|\hphi| < \alpha$ both do.
\end{enumerate}

\begin{figure}[h!]
\begin{center}
\includegraphics[width = 0.48\textwidth]{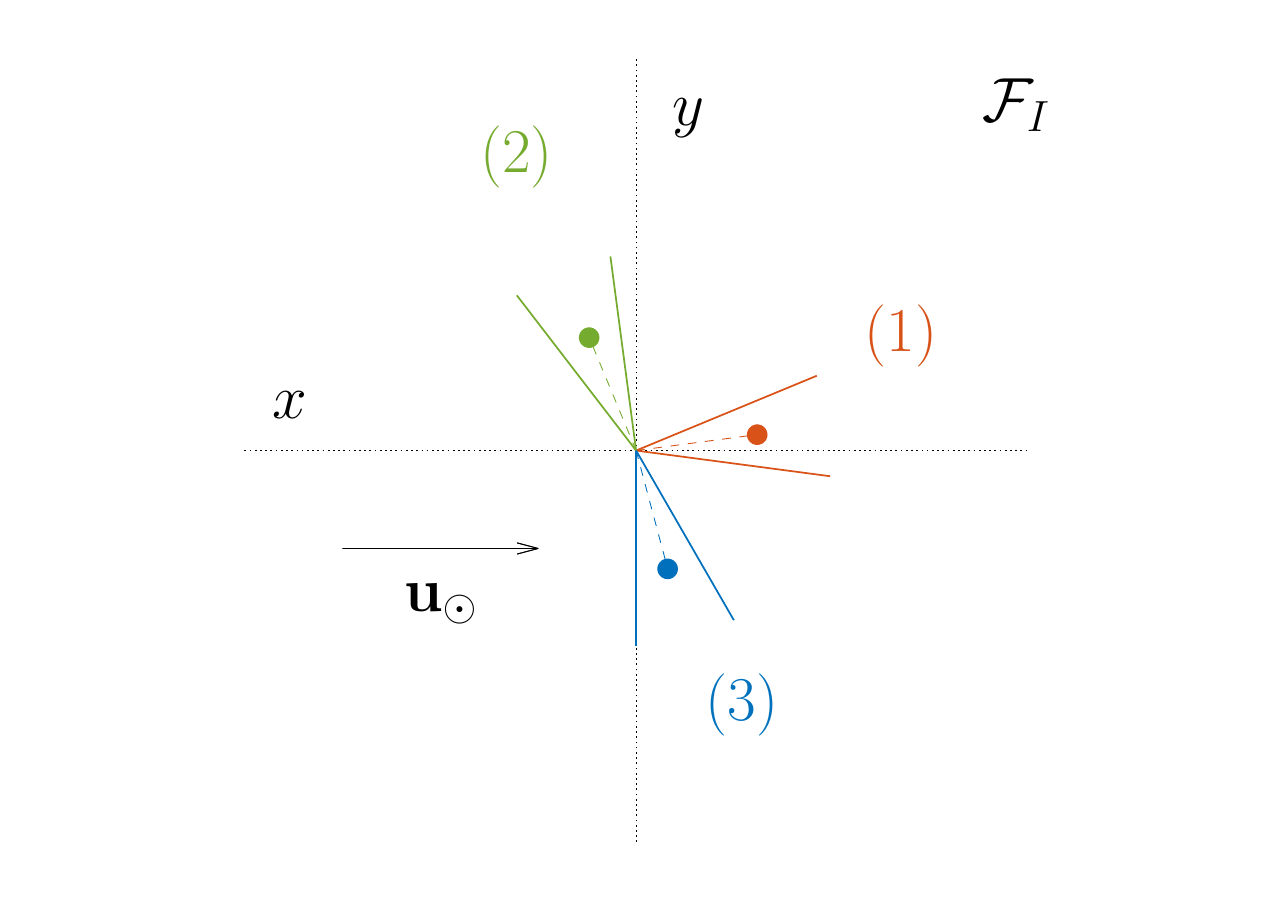}
\end{center}
\caption{Sketch of different orientations of the Sail in the $\Fcal_I$ frame in
the case of SRP, where $\lambda$ is depicted as being $180^\circ$. If oriented
as depicted, only $P_+$ (resp. $P_-$) produces torque for (2) (resp. (3)),
while both panels do in (1).} 
\label{fig:numpanels}
\end{figure}

This is sketched in Fig.~\ref{fig:numpanels}, where the three cases are
exemplified: in the sketch $\lambda=180^\circ$ and if so, both $P_\pm$ produce
torque in (1), and only $P_+$ in (3) and only $P_-$ in (2).

% - Figura

%-SRP
\subsubsection{Solar radiation pressure}\label{subsubsec:SRP}

The force due to SRP exerted in each panel of the sail is assumed to be
partially specularly reflected and partially absorbed~\cite{MC14}
\begin{eqnarray}\label{eq:forcesrp}
\bfF_{\rm SRP}^\pm & = & -p_{\rm SR}A_s (\bfn_\pm\cdot \bfu_{\rm S})
\left(
2 \eta (\bfn_\pm\cdot \bfu_{\rm S})\bfn_\pm + (1- \eta ) \bfu_{\rm S}
\right),
\end{eqnarray}
where $\eta\in(0,1)$ is the (dimensionless) reflectance of the sail and
$p_{\rm SR} = 4.56\times10^{-6}\;{\rm N}/{\rm m}^2$ is the solar pressure at 1
AU, that is considered to be constant.

On the one hand, transporting the normal vectors of the panels to $\Fcal_I$,
$R_3(\vp)\bfn_\pm$, one can see that the SRP acceleration due to the panel
$P_\pm$ reads $(a_x^\pm, a_y^\pm,0)_I^\top$, where
\begin{subequations}\label{eq:accpanelSRP}
\begin{eqnarray}
a_x^\pm & = &\frac{A_sp_{\rm SR}}{m_b+m_s}\sin(\alpha\pm\lambda\mp\vp)
        (   \eta\cos(2\alpha\pm\lambda\mp 2\vp)-\cos\lambda))\\
a_y^\pm & = &\frac{A_sp_{\rm SR}}{m_b+m_s}\sin(\alpha\pm\lambda\mp\vp)
        (\mp\eta\sin(2\alpha\pm\lambda\mp 2\vp)-\sin\lambda)).
\end{eqnarray}
\end{subequations}
Hence the acceleration due SRP can be written as
\begin{eqnarray}\label{eq:accSRP}
\frac{A_sp_{\rm SR}}{m_b+m_s}(a_x, a_y, 0)_I^\top,
\end{eqnarray}
where
\begin{eqnarray*}
(a_x, a_y, 0)_I^\top &\!\! =\!\! & \frac{m_b+m_s}{A_sp_{\rm SR}}\left[ 
(a_x^+, a_y^+, 0)_I^\top \chi_{(-\alpha,\pi-\alpha)}  (\vp-\lambda) + 
(a_x^-, a_y^-, 0)_I^\top \chi_{(-\pi + \alpha,\alpha)}(\vp-\lambda)
\right].
\end{eqnarray*}
Here 
$$
\chi_I:\Rset\to\{0,1\},
\qquad
\chi_I(\psi) = \left\{
\begin{array}{rcl}
1 & \mbox{if} & \psi\in    I\\
0 & \mbox{if} & \psi\notin I\\
\end{array}
\right.
$$
denotes the characteristic function of the interval $I$. This notation

Note that $a_{x}$ and $a_{y}$ just consist of the adimensional factors in
Eq.~\ref{eq:accpanelSRP}. In the particular case that $\alpha=\pi/2$ and $\vp =
\lambda$, that is, when the sail is a rectangular flat panel with sides $2w$
and $w$, and the direction of the normal is parallel to the sun-spacecraft
direction, then Eq.~\ref{eq:accSRP} reads
$$
-2\frac{A_sp_{\rm SR}}{m_b+m_s}(1+\eta)(\cos\lambda, \sin\lambda, 0)_I^\top,
$$
twice the acceleration of a flat square panel of size $w$ always oriented
towards the Sun\footnote{In the literature the notation $c_R=1+\eta$ is used,
and it is referred to as reflectivity coefficient.}, see~\cite{CBF16}.\\

On the other hand, the torque due to SRP is
$
\bfM_{\rm SRP} = \bfM_{\rm SRP}^- + \bfM_{\rm SRP}^+,
$
where
$\bfM_{\rm SRP}^\pm=% & = &
\bfr_\pm \times \bfF_{\rm SRP}^\pm
$, 
and $\bfr_\pm$ is the location of the center of mass of the sail panel
$P_\pm$. Their third component $M_{\rm SRP,\zeta}^\pm$ reads
\begin{eqnarray}\label{eq:SRPpanel}
M_{\rm SRP,\zeta}^\pm & = &
\frac{A_s}{m_b+m_s}\frac{p_{\rm SR}}{2}
\left(k_{1,1}(\eta)\sigma_1\sigma_2 \pm k_{2,0}(\eta)\sigma_1^2 
\pm k_{0,2}(\eta)\sigma_2^2\right),
\end{eqnarray}
where
\begin{subequations}\label{eq:coeftorque}
\begin{eqnarray}
k_{1,1}(\eta) & = & \sin\alpha\left[2 d m_b (2\eta\cos(2\alpha) + \eta + 1) + 
w (m_b + m_s)(\cos\alpha - \eta\cos(3\alpha))\right], \label{eq:a11}\\
k_{2,0}(\eta) & = & \sin^2\alpha\left[4 d \eta m_b\cos\alpha + w(m_b + m_s)(1-\eta\cos(2\alpha))\right],\mbox{ and}\label{eq:a20}\\
k_{0,2}(\eta) & = & \cos\alpha\left[2dm_b(\eta\cos(2\alpha)+1)+\eta w(m_b+m_s)\sin\alpha\sin(2\alpha)\right].\label{eq:a02}
\end{eqnarray}
\end{subequations}
Hence, using Eqs.~\ref{eq:vectors} and \ref{eq:earthsun} and assuming
that\footnote{The consequences of such a choice are explained in
\S~\ref{subsubsec:unpert}.} $k_{1,1}(\eta)\neq0$, the torque due to SRP can be
written as follows: denoting 
\begin{eqnarray}\label{eq:M0}
M_0^\pm(\psi,\eta)=-\frac{1}{2}\sin(2\psi)\pm\frac{k_{2,0}(\eta)}{k_{1,1}(\eta)}\cos^2\psi
                                     \pm\frac{k_{0,2}(\eta)}{k_{1,1}(\eta)}\sin^2\psi,
\end{eqnarray}
then
\begin{subequations}\label{eq:SRPtorque-tot}
\begin{eqnarray}\label{eq:SRPtorque}
M_{\rm SRP} & = & \frac{A_s}{m_b+m_s}\frac{p_{\rm SR}k_{1,1}(\eta)}{2} 
		M_1(\vp-\lambda, \eta),
		\quad\mbox{ where}\label{eq:SRPtorque}\\
M_1(\psi, \eta) &\!\! =\!\! & M_0^-(\psi, \eta)\chi_{[-\alpha,\pi-\alpha]}(\psi) +
                   M_0^+(\psi, \eta)\chi_{[-\pi+\alpha,\alpha]}(\psi)
			\label{eq:SRPtorque_easy}
\end{eqnarray}
\end{subequations}

The functions $M_0^\pm$ has to be understood as the scaled torque due to
$P_\pm$, and $M_1$ as the torque due to both panels at the same time, as it
takes into account the orientation of the sunlight direction with respect to
the panels.

The coefficients $k_{1,1},k_{2,0}$ and $k_{0,2}$ also depend on the masses
$m_b,m_s$, the parameters $\alpha, d$ and the width $w$, but only the
dependence on $\eta$ is stressed for reasons that are clarified in
\S~\ref{subsubsec:DRAG}, that is devoted to the equations of the effects due to
atmospheric drag. 

%-DRAG
\subsubsection{Atmospheric drag}\label{subsubsec:DRAG}

As for SRP the force due to atmospheric drag can be decomposed as the sum of
the forces exerted to each of the two panels when experiencing air
resistance. Due to the similarity with the procedure for SRP in
\S~\ref{subsubsec:SRP}, some details of the derivation of the formulas are
omitted. 

The force due to drag of each panel can be written as~\cite{MC14}
\begin{eqnarray}\label{eq:forcedrag}
\bfF_{\rm drag}^\pm & = &
   -\frac{1}{2}\rho v_{\rm rel}^2 C_D A_s (\bfn_\pm \cdot \bfu_{\rm rel})
\bfu_{\rm rel},
\end{eqnarray}
where $\rho$ is the atmospheric density and $C_D\in(1.5, 2.5)$ is an
empirically determined dimensionless drag coefficient. 

On the one hand, the atmospheric drag acceleration due to $P_\pm$ reads
$(b_x^\pm, b_y^\pm,0)_I^\top$, where
\begin{subequations}\label{eq:accpaneldrag}
\begin{eqnarray}
b_x^\pm & = &\frac{A_s\rho C_D}{m_b+m_s}
             v_x\left(-\sin(\alpha\mp\vp)v_x \mp \cos(\alpha\mp\vp)v_y\right)\\
b_y^\pm & = &\frac{A_s\rho C_D}{m_b+m_s}
             v_y\left(-\sin(\alpha\mp\vp)v_x \mp \cos(\alpha\mp\vp)v_y\right)
\end{eqnarray}
\end{subequations}
Hence the acceleration due SRP can be written as
\begin{eqnarray}\label{eq:accdrag}
\frac{A_s\rho C_D}{m_b+m_s}(b_x, b_y, 0)_I^\top,
\end{eqnarray}
where
\begin{eqnarray*}
(b_x, b_y, 0)_I^\top &\!\! =\!\! & \frac{m_b+m_s}{A_s\rho C_D}\left[ 
(b_x^+, b_y^+, 0)_I^\top \chi_{[-\alpha,\pi-\alpha]}  (\vp-\delta) + 
(b_x^-, b_y^-, 0)_I^\top \chi_{[-\pi + \alpha,\alpha]}(\vp-\delta)
\right].
\end{eqnarray*}

Note that in the particular case $\vp = \delta$ and $\alpha=\pi/2$, that is,
when the two panels form a single rectangular panel of sides $2w$ and $w$ that
is always perpendicular to the relative velocity vector, one recovers the known
formula for the drag acceleration
$$
-2\frac{A_s\rho C_D}{m_b+m_s}v_{\rm rel}^2(\cos\delta,\sin\delta,0)_I^\top.
$$

On the other hand, the torque due to this force is,
$
\bfM_{\rm drag} = \bfr_- \times \bfF_{\rm drag}^- + 
                  \bfr_+ \times \bfF_{\rm drag}^+,
$ 
whose third component reads
\begin{eqnarray}\label{eq:dragpanel}
M_{\rm drag,\zeta}^\pm & = &
\frac{A_s}{m_b+m_s}\frac{\rho v_{\rm rel}^2 C_D}{4}
\left(k_{1,1}'\nu_1\nu_2\pm k_{2,0}'\nu_1^2\pm k_{0,2}'\nu_2^2\right),
\end{eqnarray}
where
\begin{subequations}\label{eq:coeftorquedrag}
\begin{eqnarray}
k_{1,1}' & = & \sin\alpha\left[2 d m_b + w m_b \cos\alpha\right], \label{eq:b11}\\
k_{2,0}' & = & \sin^2\alpha\left[w(m_b + m_s)\right],\mbox{ and}\label{eq:b20}\\
k_{0,2}' & = & \cos\alpha\left[2dm_b\right].\label{eq:b02}
\end{eqnarray}
\end{subequations}
Note that Eqs.~\ref{eq:coeftorquedrag} and Eqs.~\ref{eq:coeftorque} are related
as follows: $k_{1,1}' = k_{1,1}(0)$, $k_{2,0}' = k_{2,0}(0)$ and $k_{0,2}' =
k_{0,2}(0)$.  Taking this into account, the torque due to atmospheric drag can
be written as
\begin{eqnarray}\label{eq:DRAGtorque}
M_{\rm drag} & = & \frac{A_s}{m_b+m_s}\frac{\rho v_{\rm rel}^2C_Dk_{1,1}(0)}{2}M_1(\vp - \delta, 0),
\end{eqnarray}
recall Eq.~\ref{eq:SRPtorque-tot}.

%-GG
\subsubsection{Gravity gradient}

The rotation of asymmetrical bodies are affected by a torque due to gravity
gradient that can be written as~\cite{MC14}
$$
\bfM_{\rm GG} = \frac{3\mu}{r_{\rm E}^3} \bfu_{\rm E}\times 
 {\bf I}_{\rm sc} \bfu_{\rm E},
$$
where $\mu=G M_{\rm E}=3.986\times10^{14}\;{\rm m}^3/{\rm s}^2$ is the
gravitational parameter of the Earth and ${\bf I}_{\rm sc} = {\rm diag}(A,B,C)$
is the inertia tensor of the spacecraft. Its component in the $\zeta$ direction is 
\begin{eqnarray}\label{eq:torquegg}
M_{\rm GG} & = & 
\frac{3 \mu}{r_{\rm sc}^3}(B-A) \gamma_1 \gamma_2
=
\frac{3 \mu}{r_{\rm sc}^3}
\left(I_{y,b} - I_{x,b} - D(\alpha, d)\right) \gamma_1 \gamma_2,
\end{eqnarray}
In practice we assume a symmetric bus, so the factor in the parenthesis of the
right hand side in Eq.~\ref{eq:torquegg} reduces to $-D(\alpha, d)$.

%-----------------------------------------------------------
% Orbit dynamics 
%-----------------------------------------------------------
\subsection{Orbit dynamics}\label{subsec:orbdyn}

As previous contributions related to the usage of the SRP perturbation for the
design of end-of-life disposals do, see~\cite{LCI12, CLI12, LCI13, CBF16}, the
$J_2$ perturbed Kepler problem is considered here as orbit dynamics. This problem is integrated in Cartesian coordinates, and the equations read
\begin{subequations}\label{eq:dyntra-dim}
\begin{eqnarray}
\ddot{x} & = & -\frac{\mu x}{r^3} - \frac{3R^2\mu J_2}{2}\frac{x}{r^5} + 
                {\rm Acc}_x\\
\ddot{y} & = & -\frac{\mu y}{r^3} - \frac{3R^2\mu J_2}{2}\frac{y}{r^5} + 
                {\rm Acc}_y
\end{eqnarray}
\end{subequations}
where $J_2=1.082\times10^{-3}$ is the adimensional $J_2$ coefficient and the
vector $({\rm Acc}_x,{\rm Acc}_y,0)_I^\top$ refers to the disturbing
acceleration under consideration, that is, either that of SRP in
Eq.~\ref{eq:accSRP} or that of atmospheric drag in Eq.~\ref{eq:accdrag}.

%-----------------------------------------------------------
% Attitude equations 
%-----------------------------------------------------------
\subsection{Attitude dynamics}\label{subsec:attdyn}

The attitude equations are those in Eq.~\ref{eq:EulerSimp}, where $M_\zeta$ is
considered to be either $M_{\rm SRP} + M_{\rm GG}$ or $M_{\rm drag} + M_{\rm
GG}$, see Eqs.~\ref{eq:SRPtorque}, \ref{eq:DRAGtorque} and \ref{eq:torquegg}.
It can be written as
\begin{eqnarray}\label{eq:dynatt-dim}
\ddot{\vp} & = & M_{\star} 
	     +  
\displaystyle    \frac{3\mu}{r_{\rm sc}^3}\frac{D(\alpha, d)}{C}
                 \sin\left(2(\arctan(y/x)-\vp)\right), 
\end{eqnarray}
where $\star$ is either SRP or drag. 

%-----------------------------------------------------------
% Values 
%-----------------------------------------------------------
\subsection{Case studies of the simulations}

There are some aspects of the dynamics that can be studied analytically and
some features can be explained via arguments of the theory of dynamical
systems. The complete system depends on many independent parameters that
account for the size, shape, mass distribution etc. of the spacecraft. Some of
these free independent parameters can be related to each other if we impose
that the spacecraft under study is feasible according to current technological
constraints. In \cite{DV18} the authors provide a way to check if, given $m_b$
and an area-to-mass ratio, it is feasible to construct a solar sail with these
requirements, and they provide a way to obtain the side-length of such a
(square) sail, and which should be its mass.

To exemplify the results of this work, two spacecraft whose sails consist of
two equal square panels as in the sketch in Fig.~\ref{fig:shapesail} are
considered, with reflectance $\eta = 0.8$, and we have obtained its
measurements by assuming the conservative values $m_b = 100\;{\rm kg}$, $w = h
= 9.20\;{\rm m}$ and $m_s = 3.60\;{\rm kg}$, that give rise to an area-to-mass
ratio $A_s/(m_b + m_s) \approx 0.82\;{\rm m^2}/{\rm kg}$.

Since the results depend strongly on the physical parameters $\alpha$ and $d$,
the dynamics of 2 structures: ${\tt SC}_1$, with $\al = 30^\circ$ and $d = 0$
m, and ${\tt SC}_2$, with $\alpha = 45^\circ$ and $d = 3.37$ m have been
studied. In the left and center panels of Fig.~\ref{fig:shapesc} sketches of
top views of ${\tt SC}_1$ and ${\tt SC}_2$ are shown, respectively. These
spacecraft are characterized by the fact that in ${\tt SC}_1$ the centres of
mass of the bus and the sail structure are at the origin, and in ${\tt SC}_2$
the bus is at the tip of the sail, recall Eq.~\ref{eq:busattip}, as the example
suggested in~\cite{CHMR13}.  In Tab.~\ref{tab:physpar} the most relevant
physical parameters of the two sails are provided, obtained by assuming a
symmetric cubic bus of side-length $1$ m.

%-------------
\begin{table}[h!]
\begin{center}
\begin{tabular}{ll|cc}
%tit
&& ${\tt SC}_1$: $\al = 30^\circ$, $d = 0$ m & 
   ${\tt SC}_2$: $\al = 45^\circ$, $d = 3.37$ m\\\hline
$A$ &[kg km$\!^2$]      & $6.74506667\times10^{-5}$ & $6.74506667\times10^{-5}$ \\ 
$B$ &[kg km$\!^2$]      & $1.05538667\times10^{-4}$ & $1.22701867\times10^{-3}$ \\ 
$C$ &[kg km$\!^2$]      & $5.47546667\times10^{-5}$ & $1.17623466\times10^{-3}$ \\ 
$k_{1,1}(0.8)$ &[kg km] & $4.12713066\times10^{-1}$ & $1.71561600\times10^{0} $ \\
$k_{2,0}(0.8)$ &[kg km] & $1.42968000\times10^{-1}$ & $8.57808000\times10^{-1}$ \\
$k_{0,2}(0.8)$ &[kg km] & $2.85936000\times10^{-1}$ & $8.57808000\times10^{-1}$ \\
$k_{1,1}(0)$   &[kg km] & $4.12713066\times10^{-1}$ & $9.53120000\times10^{-1}$ \\
$k_{2,0}(0)$   &[kg km] & $2.38279999\times10^{-1}$ & $4.76560000\times10^{-1}$ \\
$k_{0,2}(0)$   &[kg km] & $0.00000000\times10^{0} $ & $4.76560000\times10^{-1}$ 
\end{tabular}
\end{center}
\caption{Physical parameters of the two structures ${\tt SC}_1$ and ${\tt
SC}_2$.}
\label{tab:physpar}
\end{table}

%---------------------------------------------------------------------- 
% Section: SRP-dominated
%----------------------------------------------------------------------
%-----------------------------------------------------------
% Auto-stabilizing structure 
%-----------------------------------------------------------
\section{Helio-stability in SRP-dominated regions}\label{sec:SRP}

This section is devoted to analyze and study the helio-stability properties of
a solar sail as sketched in Fig.~\ref{fig:shapesail}. A qualitative description
of the coupled orbit and attitude dynamics is done in \S~\ref{subsec:simplSRP}
by studying a simplified version of the full 6D system of differential
equations Eq.~\ref{eq:dyntra-dim} and Eq.~\ref{eq:dynatt-dim}. The numerical
results of the simplified and full systems are compared in
\S~\ref{subsect:boundedSRP}. 

It is important to highlight that, as stated in \S~\ref{subsubsec:SRP}, the solar
pressure is assumed to be constant in a vicinity of the Earth, that is, the
sunlight direction is assumed to be the Sun-Earth vector (-$\bfu_{S,I}$, recall
Eq.~\ref{eq:earthsun}) and the spacecraft-Sun distance is assumed to be
constant 1 AU. This implies that the solar radiation pressure depends on the
position of the Sun (and hence on time) but not on the position of the
spacecraft. The advantage of this assumption is that in some cases the SRP
acceleration can be included in the Hamiltonian formulation of the orbit
dynamics; and the attitude dynamics can be formulated, in some limit cases,
also as a Hamiltonian system~\cite{JCFJ16,MC19}.

%-----------------------------------------------------------
% Only SRP 
%-----------------------------------------------------------
\subsection{A simplified deterministic model}\label{subsec:simplSRP}

The full problem is a continuous 6D system, so to be able to understand the
role and effect of the attitude it is convenient to study first the latter as
if it was completely uncoupled from the orbit dynamics. Rotation is known to
occur in a faster time scale than translation; namely in the present case it
can be explicitly quantified via the physical parameters of the system,
see~\cite{MC19}. In this last reference the authors provide numerical evidence
of the fact that, allowing the attitude to evolve freely, the semi-major axis
and eccentricity  of the osculating orbit remain constant in average, provided
the initial attitude is close to the sun-pointing direction.

This suggests to consider a simplified model that is obtained by considering
that the attitude does not affect the orbit dynamics, that occurs on a fixed
Keplerian orbit. Assume such orbit to have fixed semi-major axis $a$,
eccentricity $e$ and argument of the perigee $\omega$. Denote the true anomaly
as $\theta$. The equations of this problem are obtained by taking
Eq.~\ref{eq:dynatt-dim} for SRP, written in Keplerian elements instead of in
Cartesian coordinates, that reads
\begin{subequations}\label{eq:dynatt-dyn-simpl}
\begin{eqnarray}
\dot{\lambda} & = & n_\odot,\label{eq:dlambda}\\
\dot{M}       & = & n,\label{eq:dM}\\
\dot{\vp}     & = & \Phi,\label{eq:dPhi}\\
\dot{\Phi}    & = & \frac{A_s}{m_b+m_s}\frac{p_{\rm SR}k_{1,1}(\eta)}{2C} 
                M_1(\vp-\lambda, \eta) 
             -  
		    \frac{3\mu}{r_{\rm sc}^3}\frac{D(\alpha, d)}{C}
                 \sin\left(2(\theta+\omega-\vp)\right),\label{eq:dvp}
\end{eqnarray}
\end{subequations}
where $M$ is the mean anomaly and $n=\sqrt{\mu/a^3}$ is the mean motion.

The shape of the sail structure is chosen to ``follow" the Sun along its
apparent orbit, and this fact suggests to change variables as follows:
$$
\tilde\lambda = \lambda,    \quad
\tilde{M} = M,              \quad
\tilde{\vp} = \vp - \lambda \quad
\mbox{ and }                \quad
\tilde\Phi = \Phi - n_\odot.
$$ 

To obtain a set of equations with which one can have a unified understanding of
the dynamics it is convenient to adimensionalise time by choosing
\begin{eqnarray}\label{eq:timescaling}
t = t_\star\;s, \qquad
t_\star^2 = \frac{m_b + m_s}{A_s}\frac{2C}{k_{1,1}(\eta)p_{\rm SR}}.
\end{eqnarray}
Note that the time scale $s$ is faster than the original $t$, and hence
$t_\star$ can be understood as a measure of the difference between the
different characteristic time scales of the attitude and the orbit. As already
assumed when defining Eq.~\ref{eq:M0} in \S~\ref{subsubsec:SRP}, the time
scaling makes sense only if $k_{1,1}(\eta)\neq0$, and this coefficient vanishes
only at
\begin{eqnarray}\label{eq:reld-al}
d = \frac{w(m_b+m_s)}{2m_b}K(\alpha,\eta),\quad
\mbox{where}\quad
K(\alpha,\eta) = \frac{\eta\cos(3\alpha)-\cos\alpha}
                      {2\eta\cos(2\alpha)+\eta+1}.
\end{eqnarray}
Moreover, since $\eta\in(0,1)$ and $\alpha\in(0,\pi/2)$, $K(\alpha,\eta) < 0$.

Note that $t_\star^2$ is the inverse of the prefactor of $M_1$ in
Eq.~\ref{eq:dvp}, and this scaling depends solely on physical parameters of the
system, namely on the size and mass distribution. A final change of variables
that separates slow and fast components 
$$
\hat{\lambda} = \tilde{\lambda},    \quad
\hat{M} = \tilde{M},              \quad
\hat{\vp} = \tilde{\vp} \quad
\mbox{ and }                \quad
\hat{\Phi} = t_\star \tilde{\Phi},
$$
puts the equations in the final form in which they will be dealt with. Let
$\hat{\lambda}_0$ denote an initial value of $\hat{\lambda}$. The
equations read
\begin{subequations}\label{eq:dynatt-dyn-simpl-fin}
\begin{eqnarray}
\hat{\lambda}  & = & t_\star n_\odot \label{eq:dl-fin}\\
\hat{M}'       & = & t_\star n,\label{eq:dM-fin}\\
\hat{\vp}'     & = & \hat{\Phi},\label{eq:dPhi-fin}\\
\hat{\Phi}'    & = & M_1(\hat{\vp}, \eta) +  
                    t_\star^2\frac{3\mu}{r_{\rm sc}^3}\frac{D(\alpha, d)}{C}
\sin\left(2(\theta+(\omega - \hat{\lambda})-\hat{\vp})\right),\label{eq:dvp-fin}
\end{eqnarray}
\end{subequations}
where the derivative with respect to $s$ is indicated with $(\;')={\rm d}/{\rm
d}s$. Written like this, formally, Eq.~\ref{eq:dM-fin} is slow and
Eqs.~\ref{eq:dPhi-fin} and \ref{eq:dvp-fin} are fast.

The vector field Eq.~\ref{eq:dynatt-dyn-simpl-fin} is continuous but not
differentiable at $\hat{\vp} = \pm\alpha$ because $M_1$ is piecewise defined,
recall Eq.~\ref{eq:SRPtorque_easy}; yet it is Lipschitz continuous, so the
theorem of existence and uniqueness of solutions of Ordinary Differential
Equations (ODE) applies.\\

Still after reducing the orbit dynamics to occur on a fixed Keplerian ellipse
the problem is far from being trivial. Notice that Eq.~\ref{eq:dvp-fin} has two
summands, and the second has $t_\star^2$ as prefactor; which has, in turn, the
inverse of the area-to-mass ratio of the spacecraft as factor, see
Eq.~\ref{eq:timescaling}. This suggests to interpret
Eq.~\ref{eq:dynatt-dyn-simpl-fin} as a perturbation problem, where the gravity
gradient torque is a perturbing effect.

%-----------------------------------------------------------
% Unperturbed motion 
%-----------------------------------------------------------
\subsubsection{Dynamics neglecting gravity gradient
perturbation}\label{subsubsec:unpert}

Consider Eq.~\ref{eq:dynatt-dyn-simpl-fin} by setting $t_\star = 0$, that is,
without the gravity gradient effect. This situation can be physically
interpreted as having an arbitrarily large area-to-mass ratio. Since $M_1$ does
not depend neither on $\theta$ nor on $\lambda$, the problem reduces to
$$
\dot{\hat{\vp}}  = \hat{\Phi},\quad
\dot{\hat{\Phi}} = M_1(\hat{\vp},\eta).
$$
This system was studied in~\cite{MC19}, where it was found to have Hamiltonian
structure. Namely, if one defines
\begin{eqnarray}\label{eq:hamfast}
\begin{array}{rcl}
\displaystyle
\Kcal(\hat{\vp},\hat{\Phi},\eta) = \frac{\hat{\Phi}^2}{2}
 & + &
 K_1(\hat{\vp},\eta) \chi_{[-\alpha,\pi-\alpha]}(\hat{\phi})
+K_1(-\hat{\vp},\eta)\chi_{[-\pi+\alpha,\alpha]}(\hat{\phi})\\
 & + &
 K_1(\pi-\alpha,\eta)\chi_{[-\pi,-\pi+\alpha)\cup(\pi-\alpha,\pi]}(\hat{\vp}),
\end{array}
\end{eqnarray}
where
\begin{eqnarray*}
K_0(\psi,\eta) & = & - 
\left[\frac{1}{4}\cos(2\psi) +
\frac{1}{4k_{1,1}(\eta)}
\left((k_{2,0}(\eta)-k_{0,2}(\eta))\sin(2\psi) +
     2(k_{2,0}(\eta)+k_{0,2}(\eta))\psi
\right)
\right],\\
K_1(\psi,\eta) & = & K_0(\psi,\eta) - K_0(-\alpha,\eta),
\end{eqnarray*}
one can see, using trigonometric formulas for double angles, that 
\begin{eqnarray}\label{eq:nonpert-ham}
\dot{\hat{\vp}}  = \frac{\partial}{\partial
\hat{\Phi}}\Kcal(\hat{\vp},\hat{\Phi},\eta),
\quad 
\dot{\hat{\Phi}} = -\frac{\partial}{\partial
\hat{\vp}}\Kcal(\hat{\vp},\hat{\Phi},\eta).
\end{eqnarray}

\begin{figure}[h!]
\begin{center}
\includegraphics[width = 0.45\textwidth]{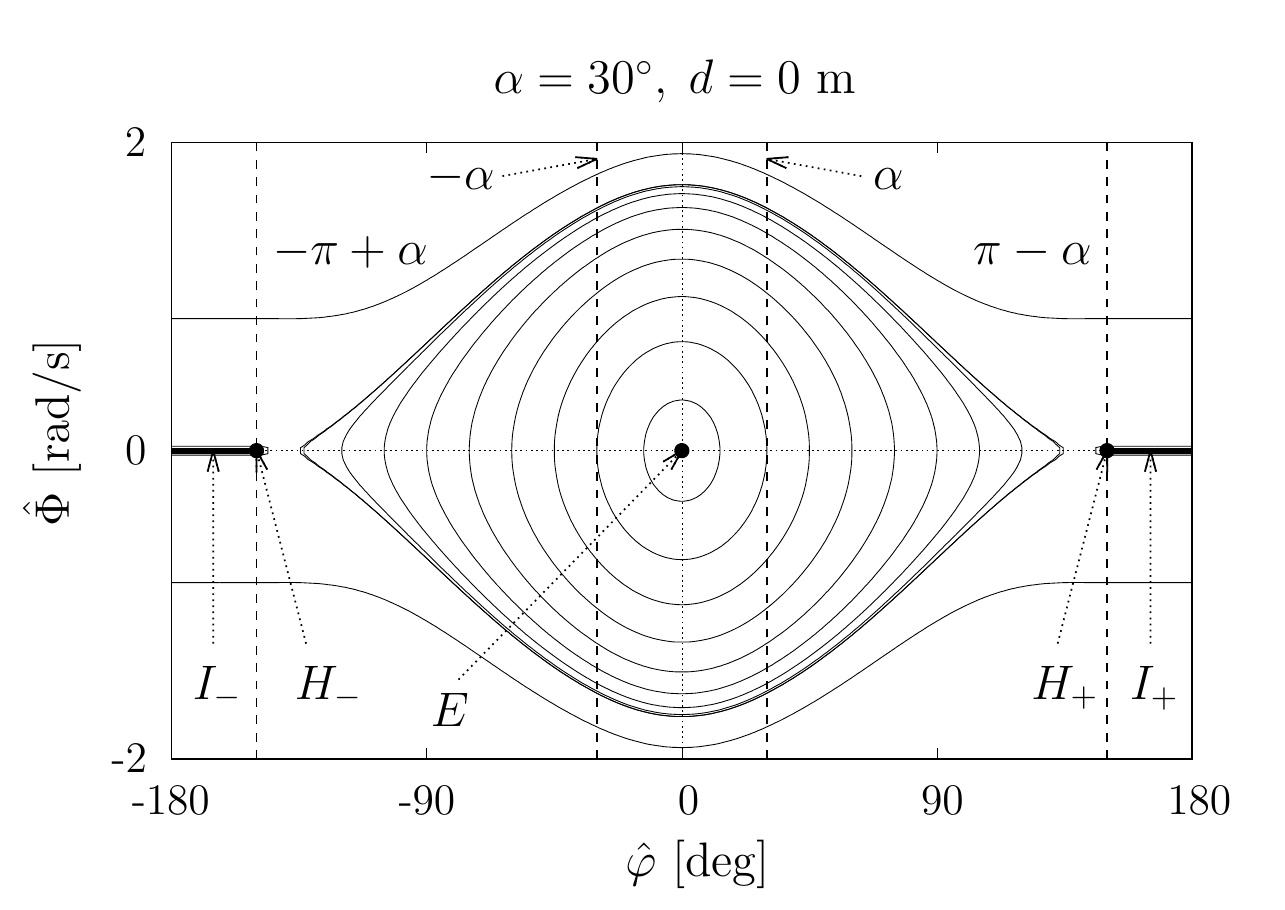}
\includegraphics[width = 0.45\textwidth]{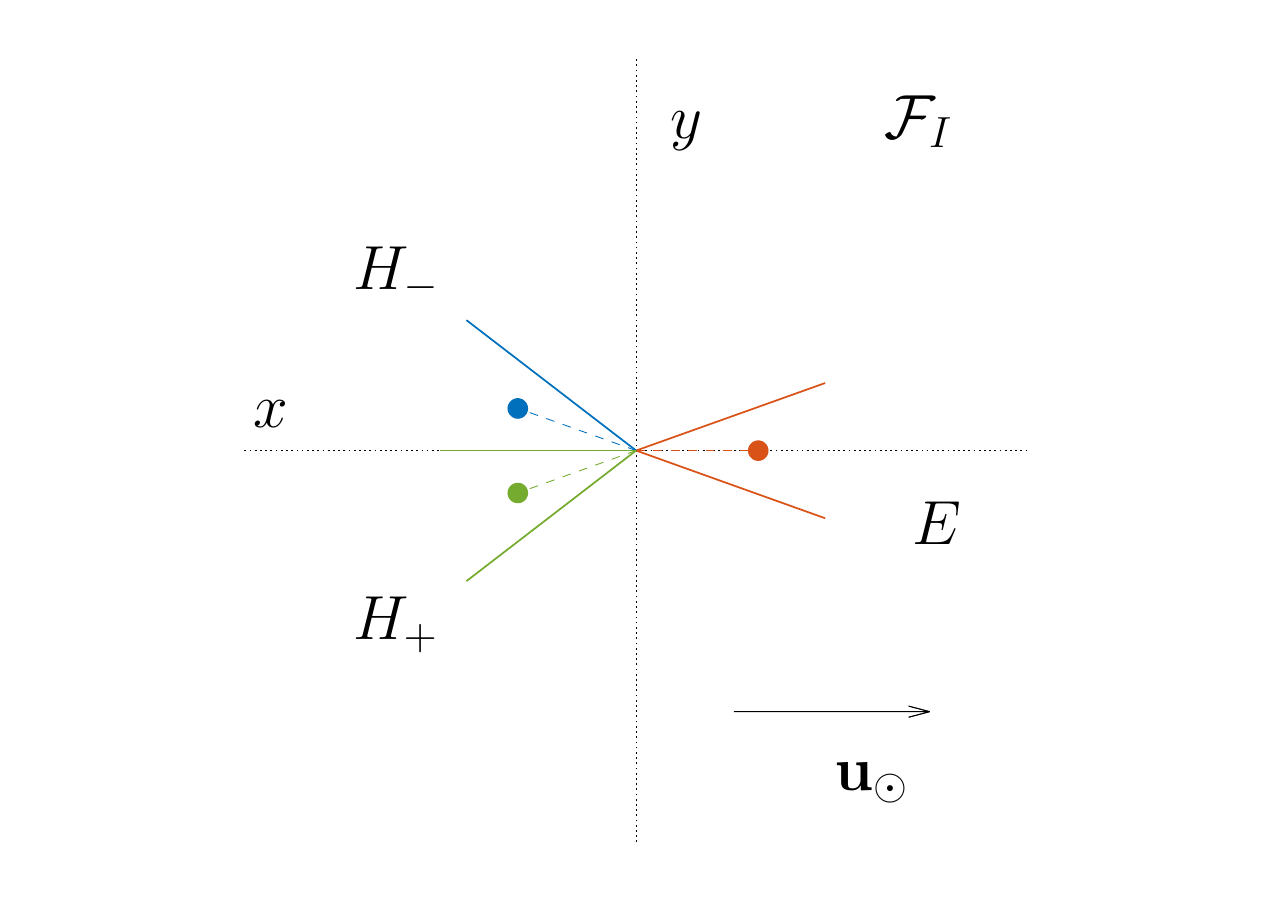}
\end{center}
\caption{Relevant orbits of Eq.~\ref{eq:nonpert-ham}. Left: Phase
space, switching manifolds (vertical dashed lines) and equilibria. Right:
Sketch of equilibrium orientations of the sail.}
\label{fig:phaseSRP}
\end{figure}

The system Eq.~\ref{eq:nonpert-ham} is $2\pi$-periodic with respect to
$\hat{\phi}$ and symmetric with respect to $\hat{\phi}=0$. The equilibria
consist of a continuum
$$
\mathcal{E} = \left\{
(\hat{\phi},0);\;
\hat{\phi}\in I_-\cup \{0\}\cup I_+,
\mbox{ where } I_- = [-\pi, -\pi + \alpha]
\mbox{ and }   I_+ = [\pi - \alpha, \pi)
\right\},
$$
plus $E = (0,0)$ that is stable provided the coefficient $k_{1,1}>0$, and
$H_\pm = (\pm \pi \mp \alpha,0)$, that are saddles whose invariant manifolds
coincide, $W^u(H_+) = W^s(H_-)$ and $W^u(H_-) = W^s(H_+)$. Note that all
equilibria in $\mathcal{E}$ do not exist when the gravity gradient torque is
added.  The phase space is depicted in the left panel of
Fig.~\ref{fig:phaseSRP}, where the equilibria and switching manifolds (where a
panel ceases or starts producing torque and, hence, differentiability is lost)
are indicated, and in the right panel the equilibria $E, H_\pm$ are sketched.

% Figura

In Fig.~\ref{fig:phaseSRP} we can see that the dynamics resembles that of a
pendulum, where orbits librate around the sun-pointing orientation $E$.
Concerning the stability of the equilibria, $H_\pm$ are unstable regardless of
the values of the parameter. On the other hand, the the sun-pointing attitude
$E$ is stable only if 
\begin{eqnarray}\label{eq:stabSRP}
k_{1,1}(\eta)>0 \quad 
\Leftrightarrow \quad
d > d_{\rm min} := \frac{w(m_b+m_s)}{2m_b}K(\alpha,\eta),
\end{eqnarray}
see Eq.~\ref{eq:reld-al}. This has to be understood as a necessary condition
for the stability of the sun-pointing attitude. Hence, we have justified the
following result.

\begin{prop} For each aperture angle $\alpha\in (0,\pi/2)$ and $\eta\in(0,1)$,
if $d> d_{\rm min}$ as defined in Eq.~\ref{eq:stabSRP} the sun-pointing
direction $E$ is a locally stable equilibrium point of
Eq.~\ref{eq:nonpert-ham}.
\end{prop} 

\begin{figure}[h!]
\begin{center}
\includegraphics[width = 0.48\textwidth]{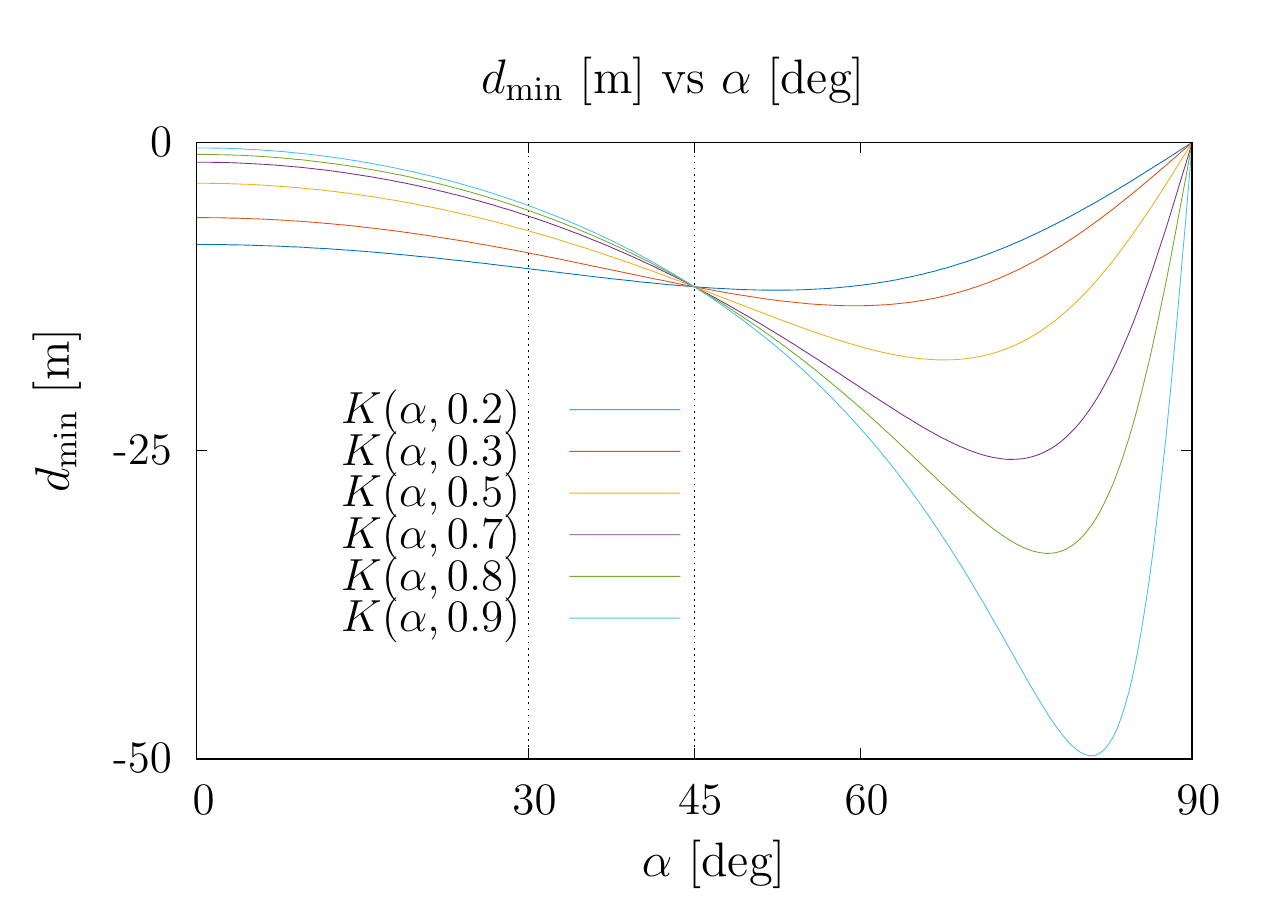}
\end{center}
\caption{Depiction of the necessary condition for the stability of the
sun-pointing attitude.}
\label{fig:dmin}
\end{figure}

Note that as $K(\alpha,\eta)<0$, $d_{\rm min}<0$ and the attitude $E$ is
stable, in particular, for all values $d>0$ and hence for all positions of the
bus in front of the sail (e.g. all three sketches in Fig.~\ref{fig:shapesc}),
even beyond the tip of the sail. The value of $d_{\rm min}$ for
$\alpha\in(0,\pi/2)$ is depicted in Fig.~\ref{fig:dmin}.

% - Figura

The motion restricted to $|\hat{\vp}|<\alpha$, that physically corresponds to
the case where both panels face sunlight (see Fig.~\ref{fig:phaseSRP} left,
between the vertical dotted lines), is that of a mathematical pendulum. Even
though Eq.~\ref{eq:nonpert-ham} is only continuous, restricted to
$|\hat{\vp}|<\alpha$ the vector field is analytic. Thanks to this property one
expects oscillations in this regime to persist for stronger gravity gradient
effects, that is, for orbits that get closer to Earth surface, as classical
averaging results apply in this region of the phase space, see~\ref{MC16}.

%-----------------------------------------------------------
% Perturbed motion 
%-----------------------------------------------------------
\subsubsection{Perturbation by gravity gradient}

Here Eq.~\ref{eq:dynatt-dyn-simpl-fin} without any assumption on $t_\star$ is
considered. Recall that this consists of adding at the same time 
\begin{enumerate}
	\item The effect of the asymmetry of the body, that is a periodic
behaviour whose period is that of the motion around the Earth, and
	\item The period of rotation of the apparent motion of the Sun around
the Earth.
\end{enumerate}
In this situation, the invariant objects (equilibria, periodic orbits) of the
system studied above in \S~\ref{subsubsec:unpert} have their dynamically
equivalent analogue in the 4D phase space of Eq.~\ref{eq:dynatt-dyn-simpl-fin}
under the conditions of smallness of $t_\star$, non-degeneracy and non-resonant
conditions. These invariant objects have to be understood as if they``gained"
the non-resonant frequencies of the perturbation \cite{JV97}: e.g. under these
hypotheses, the fixed point $E$ can have up to a 2-torus as dynamical
substitute, and libration curve in Fig.~\ref{fig:phaseSRP} can have a 3-torus
as dynamical substitute. But in this setting the existence of these objects can
only be theoretically approached in $|\hat{\vp}|<\alpha$ where the sail is
equivalent to a pendulum and hence the vector field is analytic. 

The relevance of these 3-tori is that as the phase space is 4D, these
objects separate space and hence in case they exist they define a region where
oscillations are perpetually bounded, and this orbits are strong candidates for
practically bounded orbits in the complete system. 

It is important to note that some of this structure is also expected to be
destroyed due to the gravity gradient, that is, initial conditions in
librational motion in the problem without gravity gradient can become
eventually rotational once this effect is added.  If the sail that is initially
in $|\hat{\vp}|<\pi-\alpha$ reaches $|\hat{\vp}|>\pi-\alpha$ with nonzero
angular velocity, this rotational state will never be lost. These kind of
orbits are referred to as eventually {\sl tumbling}. From a practical point of
view, this defines an {\sl escaping criterion} for simulations: the trajectory
of an initial condition is discarded as tumbling if it reaches a state
$|\hat{\vp}|>\pi-\alpha$ with nonzero angular velocity before some prescribed
integration time.\\

The smallness of the perturbation is key to be able to ensure the persistence
of tori and hence the existence of stable attitude dynamics. Since we are
assuming fixed Keplerian motion, we can write
$$
r_{\rm sc} = \frac{a\sqrt{1-e^2}}{1+e\cos \theta},
$$
Using this relation, the rightmost summand in Eq.~\ref{eq:dvp-fin} in Keplerian
elements reads
\begin{eqnarray}\label{eq:pertor}
P(\alpha, d) \frac{1}{a^3\sqrt{(1-e^2)^3}} \gamma_1\gamma_2(1 + e\cos \theta)^3,
\quad
\mbox{where}\quad
P(\alpha, d) = 3 \mu t_\star^2\frac{D(\alpha,d)}{C}.
\end{eqnarray}

\begin{figure}[h!]
\begin{center}
\includegraphics[width = 0.48\textwidth]{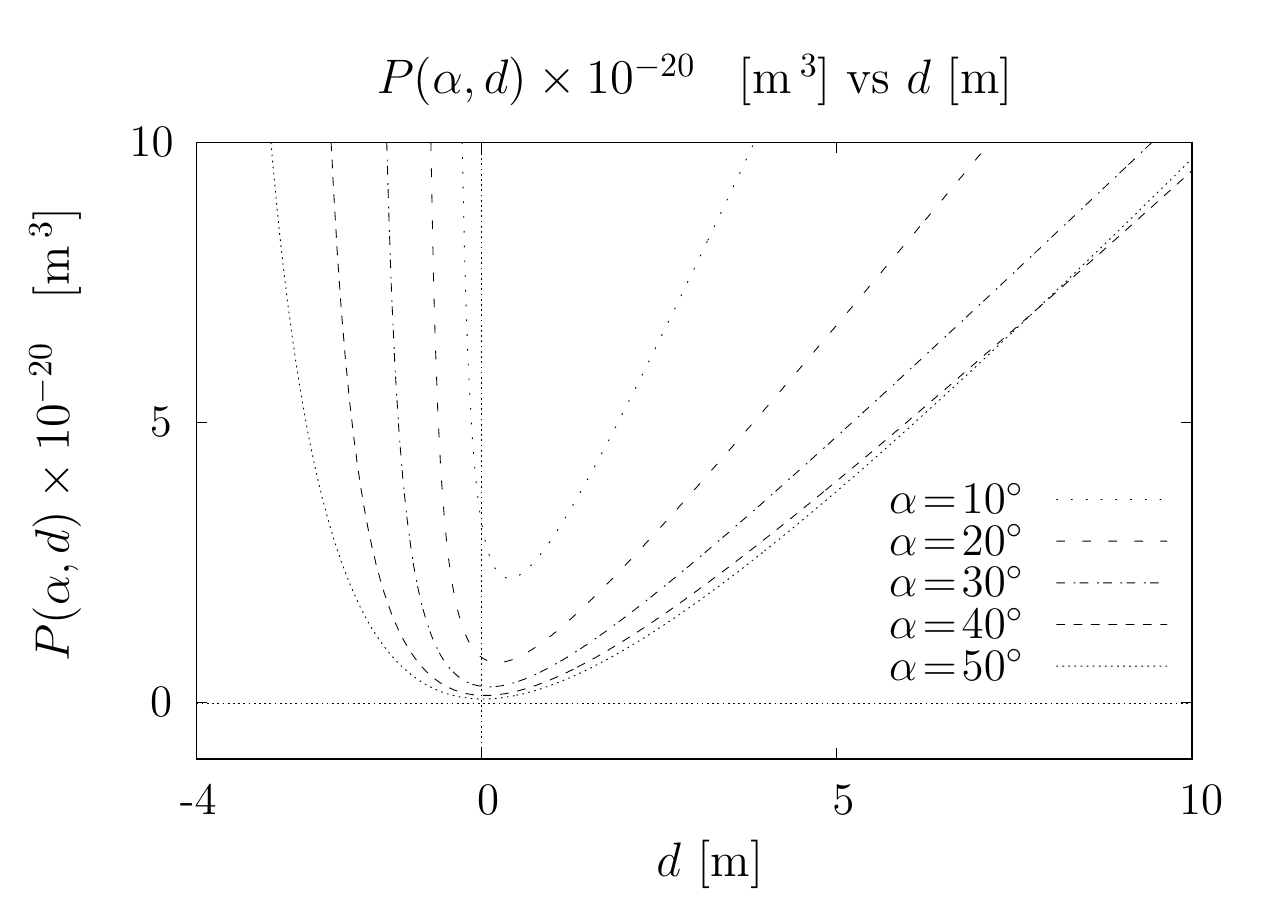}
\includegraphics[width = 0.48\textwidth]{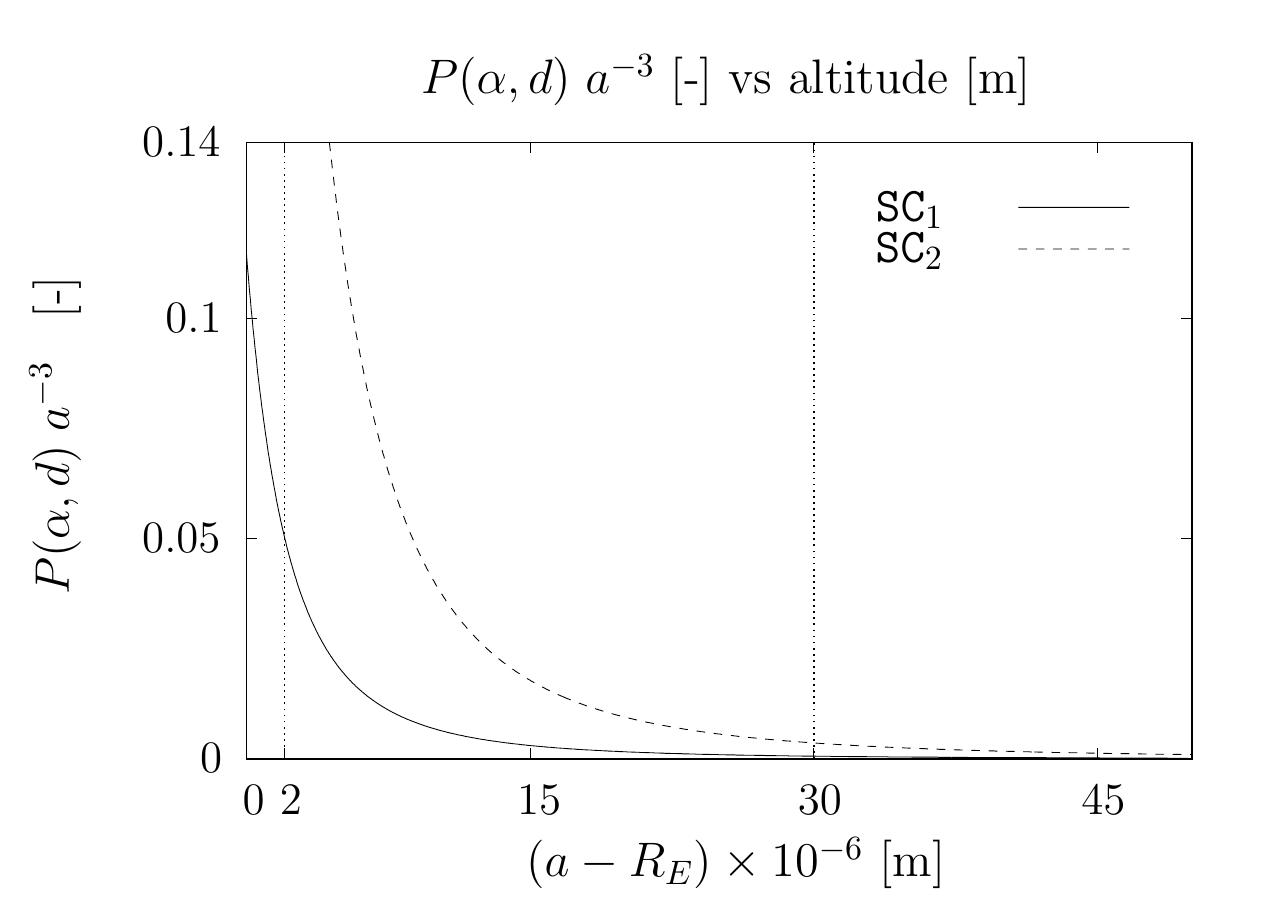}
\end{center}
\caption{Left: Size of the perturbation, $P(\alpha,d)$, see
Eq.~\ref{eq:pertor}. Here we display curves for fixed $\alpha = 10,20,30,40$
and $50$ deg. In the bottom left corner we indicate the corresponding value of
our test example. Right: $P(30,0)/(R_E + a')^3$, where $a'$ is the altitude of
the orbit.}
\label{fig:sizepert}
\end{figure}

How small the perturbation is depends on the values of $\alpha$ and $d$ via the
factor $P(\alpha, d)$ in Eq.~\ref{eq:pertor}. The full dependence on these
parameters is contained in the quotient $D(\alpha,d)/k_{1,1}(\eta)$, and it is
easy to see that, for the values where $k_{1,1}(\eta)>0$, that is, when the
orientation of the spacecraft towards the Sun is stable, for a fixed value of
$\alpha$, as a function of $d$, the quotient (and hence $P(\alpha, d)$) is
strictly convex and has an absolute minimum. In Fig.~\ref{fig:sizepert}, left,
we display some examples of $P(\alpha,d)$ for fixed values of $\alpha$.
%L'argument es fonamenta en què D/a11 és de la forma (A+Bx^2)/(C+Dx), tots
%positius menys C. La segona derivada és estrictament positiva si el
%denominador és positiu.

% - Figura

In the right panel of Fig.~\ref{fig:sizepert} we display how does the
perturbation size depend on the altitude of the orbit for ${\tt SC}_1$ and
${\tt SC}_2$. The vertical dashed lines indicate the separations from Low Earth
Orbits (LEO) and Medium Earth Orbits (MEO) at $a'=5000$ km, and between MEO and
the geostationary region (GEO) at $a'=35000$ km.\\

This shows that one has control on the size of the perturbation relying only on
the physical parameters of the system, namely on the aperture angle $\alpha$
and the MPO $d$.

%-----------------------------------------------------------
% Stable motion
%-----------------------------------------------------------
\subsection{Bounded attitude motion: simplified versus complete
model}\label{subsect:boundedSRP}

Consider first the simplified model in Eq.~\ref{eq:dynatt-dyn-simpl-fin}. Under
the presence of the gravity gradient torque, the dynamical objects of our
interest are ideally replaced by tori of dimension and 3. These orbits appear,
in a $\hat{\vp}-\hat{\Phi}$ plot, as invariant curves around $E$. If such an
orbit is detected, all initial conditions inside are attitude bounded.  If
instead one considers the attitude to affect the orbit, that is, to consider
Eq.~\ref{eq:dyntra-dim} instead of the simplification $\hat{M}'=t_\star n$
(Eq.~\ref{eq:dM-fin}) the formulation does not even provide a guess whether
there would be or not any dynamical substitute, and in case there are, what is
their dimension in the 6D phase space.  

The purpose of this section is to provide numerical evidence of the existence
of initial conditions that do not tumble before some large amount of time in
the simplified model Eq.~\ref{eq:dynatt-dyn-simpl-fin}, and that some of this
structure is actually seen in the full integration of Eq.~\ref{eq:dyntra-dim}.

\subsubsection{Numerical experiment}\label{subsubsec:numexpSRP} 
Since the orbit dynamics is always transversal to the $x$ and $y$ axes, to
detect such non-tumbling motion it is convenient to consider the hence well
defined Poincar\'e section of Eq.~\ref{eq:dynatt-dyn-simpl-fin} in 
\begin{eqnarray}\label{eq:defsigma}
\Sigma = \{x = 0,\; y < 0\},
\end{eqnarray}
that is, in the negative $y$ axis, recall the sketch in
Fig.~\ref{fig:sketchSigma}. This reduces the problem to discrete and the
dimension of the phase space by 1. 

As the scope of this paper is the study of practically stable attitude
dynamics, two orbit initial conditions are chosen: initially, ${\tt SC}_{1,2}$
are on an orbit with altitude $h_0=5000$ km, $\omega_0=0^\circ$ and $e=0.001$
or $0.1$. These are chosen to test the effect of stronger gravity gradient
effects on the same spacecraft. An equispaced grid of 40 initial attitudes is
chosen as follows
\begin{itemize}
	\item[-] For ${\tt SC}_1$, $\Phi_0 = n_\odot$ rad/s and $\vp_0 =
\lambda_0 + 0.9\times (180^\circ-30^\circ)^\circ\times j / 40$, $j = 0,\ldots,
39$, 
	\item[-] For ${\tt SC}_2$, $\Phi_0 = n_\odot + 2\times0.9\times j / 40$ rad/s,
$j = 0,\ldots, 39$, $\vp_0 = \lambda_0$.
\end{itemize}
This segments have been chosen so that the most relevant parts of the phase
space were visible in the $\hat{\vp}-\hat{\Phi}$ plot. All initial conditions
have been integrated for at most 1 year, and the integration was stopped before
only in case along the orbit $|\hat{\vp}|>\pi-\al$, that is, if the spacecraft
started tumbling.

\subsubsection{Numerical results}

\begin{figure}[h!]
\begin{center}
\includegraphics[width = 0.48\textwidth] {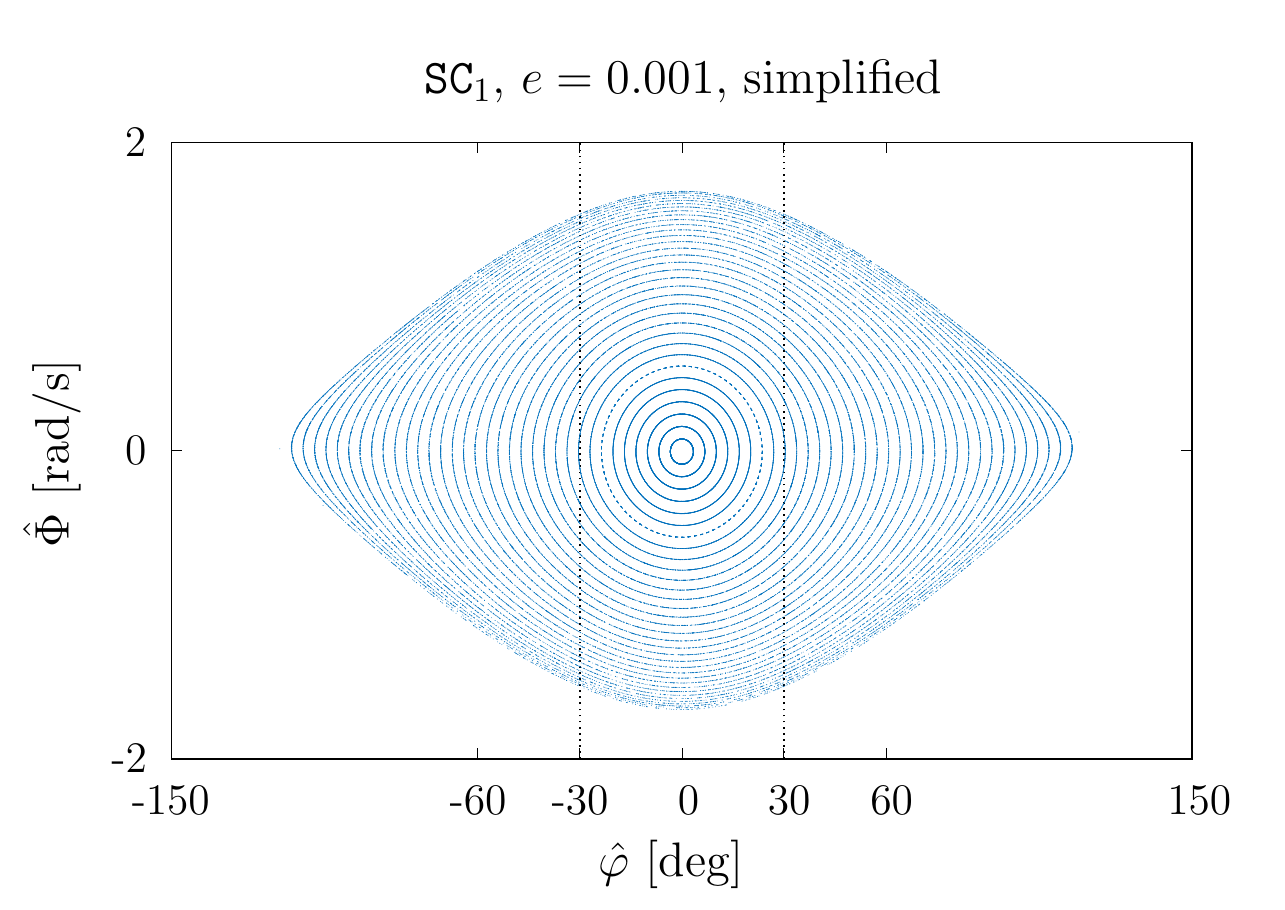}
\includegraphics[width = 0.48\textwidth] {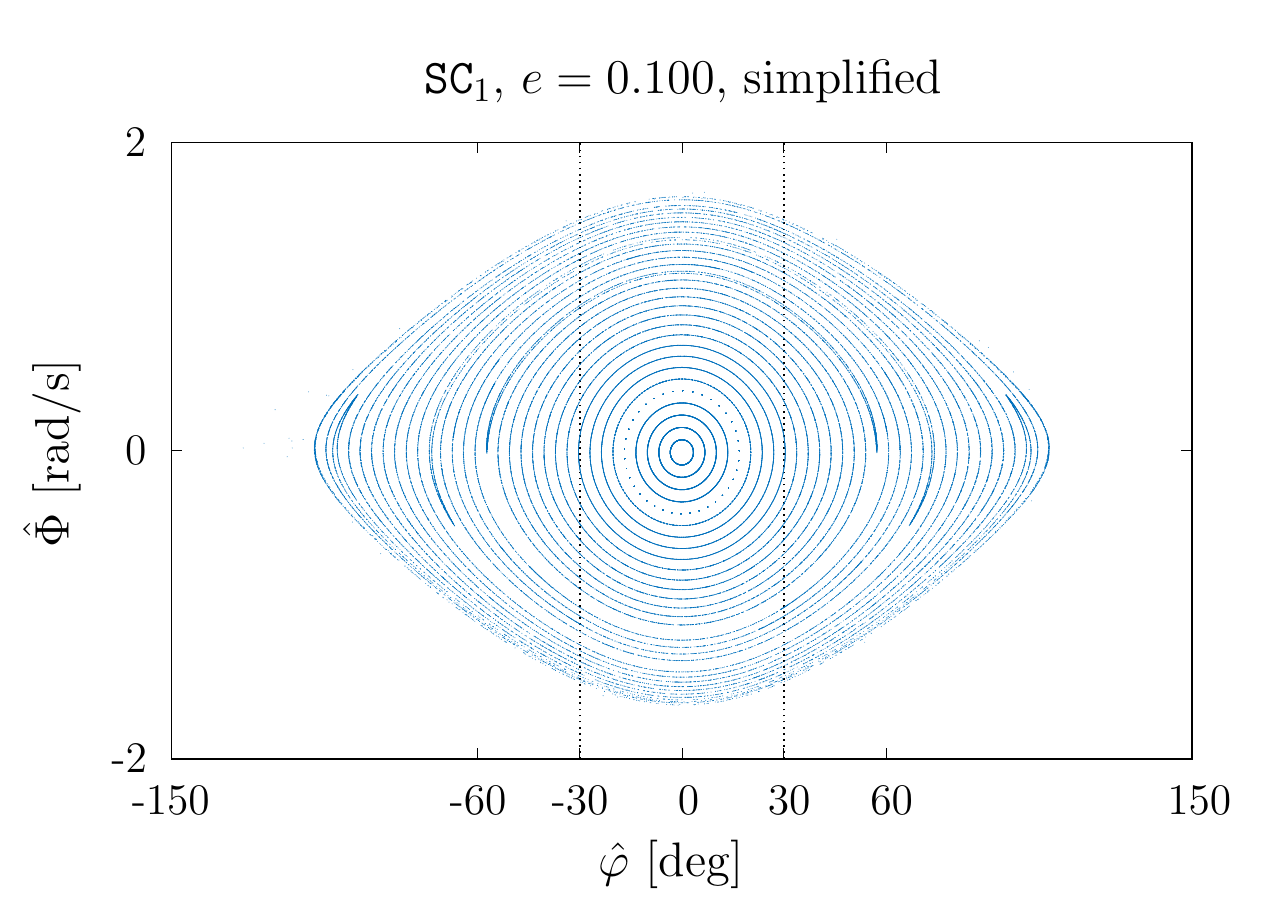}\\
\includegraphics[width = 0.48\textwidth] {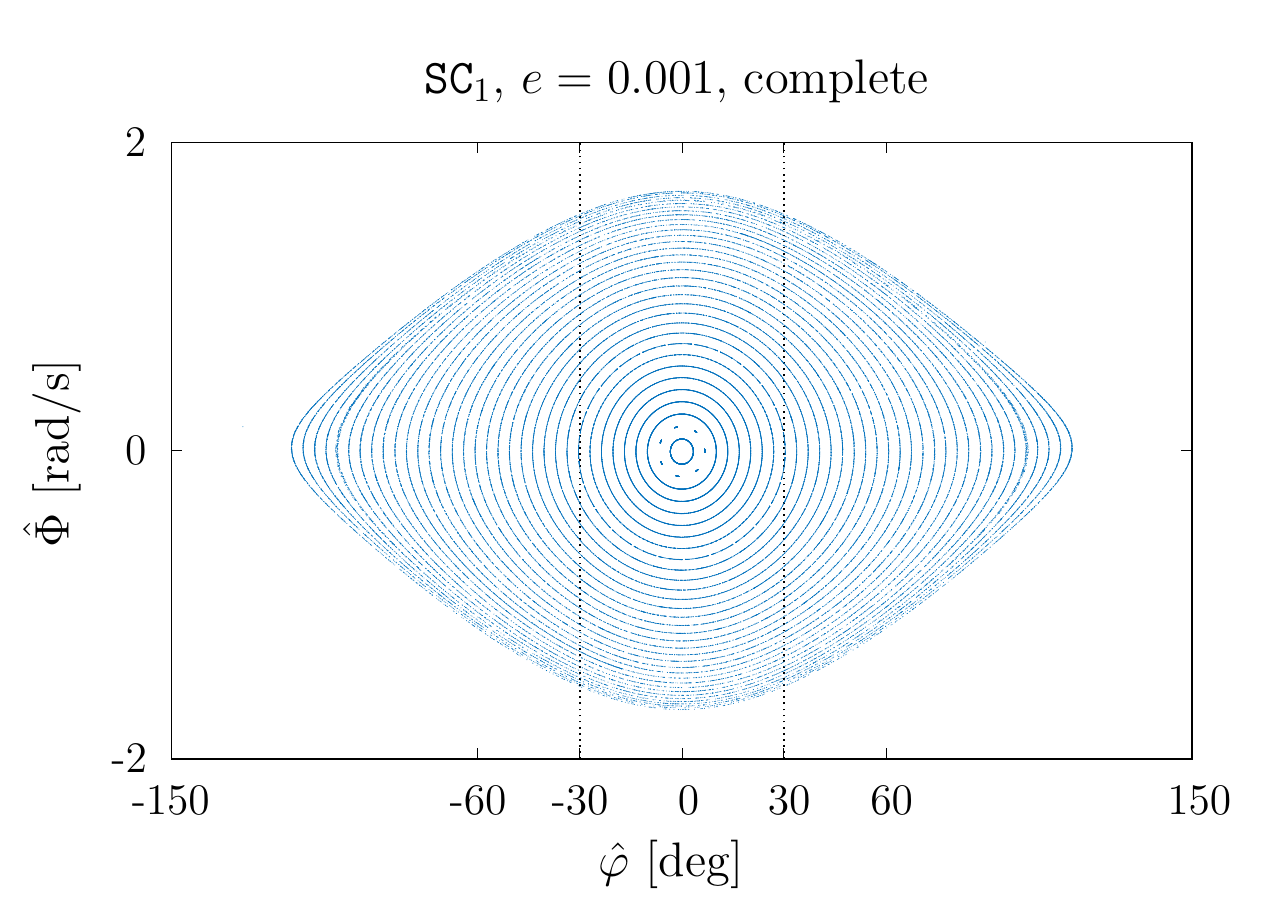}
\includegraphics[width = 0.48\textwidth] {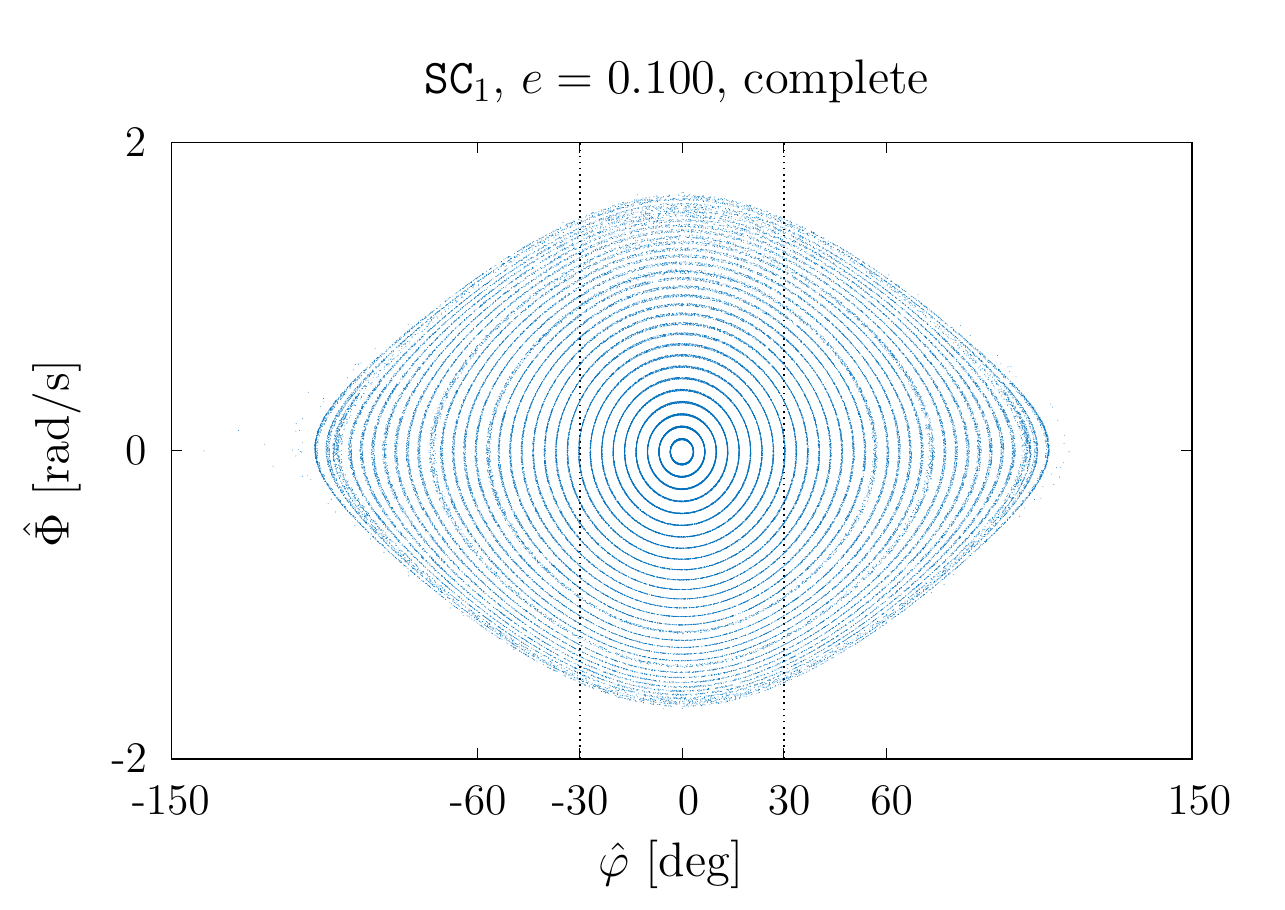}
\end{center}
\caption{Iterates of the Poincar\'e map on $\Sigma$ of the initial condition
described in \S~\ref{subsubsec:numexpSRP} for ${\tt SC}_1$. Top: simplified
model. Bottom: complete model. Left: $e=0.001$. Right: $e=0.100$.}
\label{fig:SC1-SRP}
\end{figure}

\begin{figure}[h!]
\begin{center}
\includegraphics[width = 0.48\textwidth] {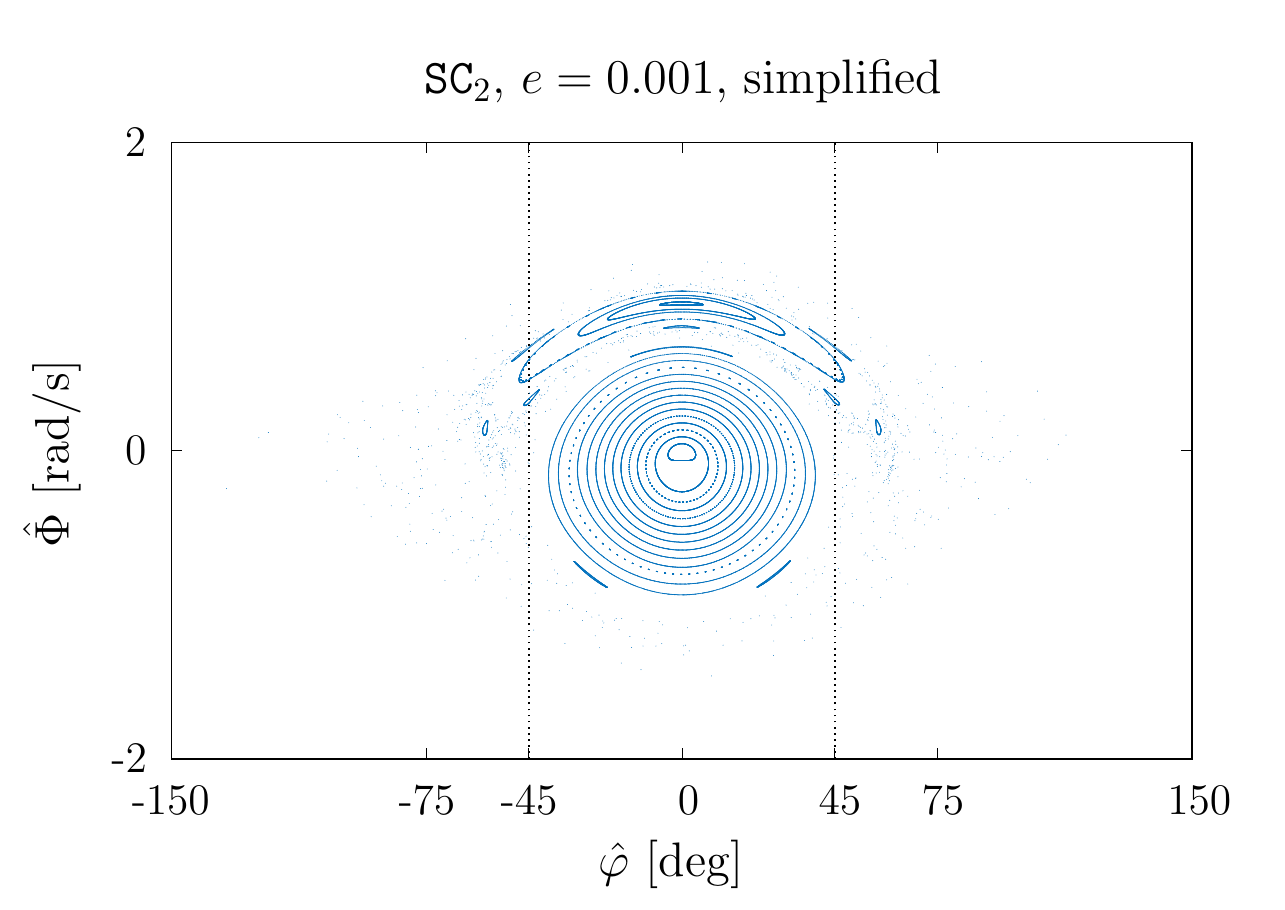}
\includegraphics[width = 0.48\textwidth] {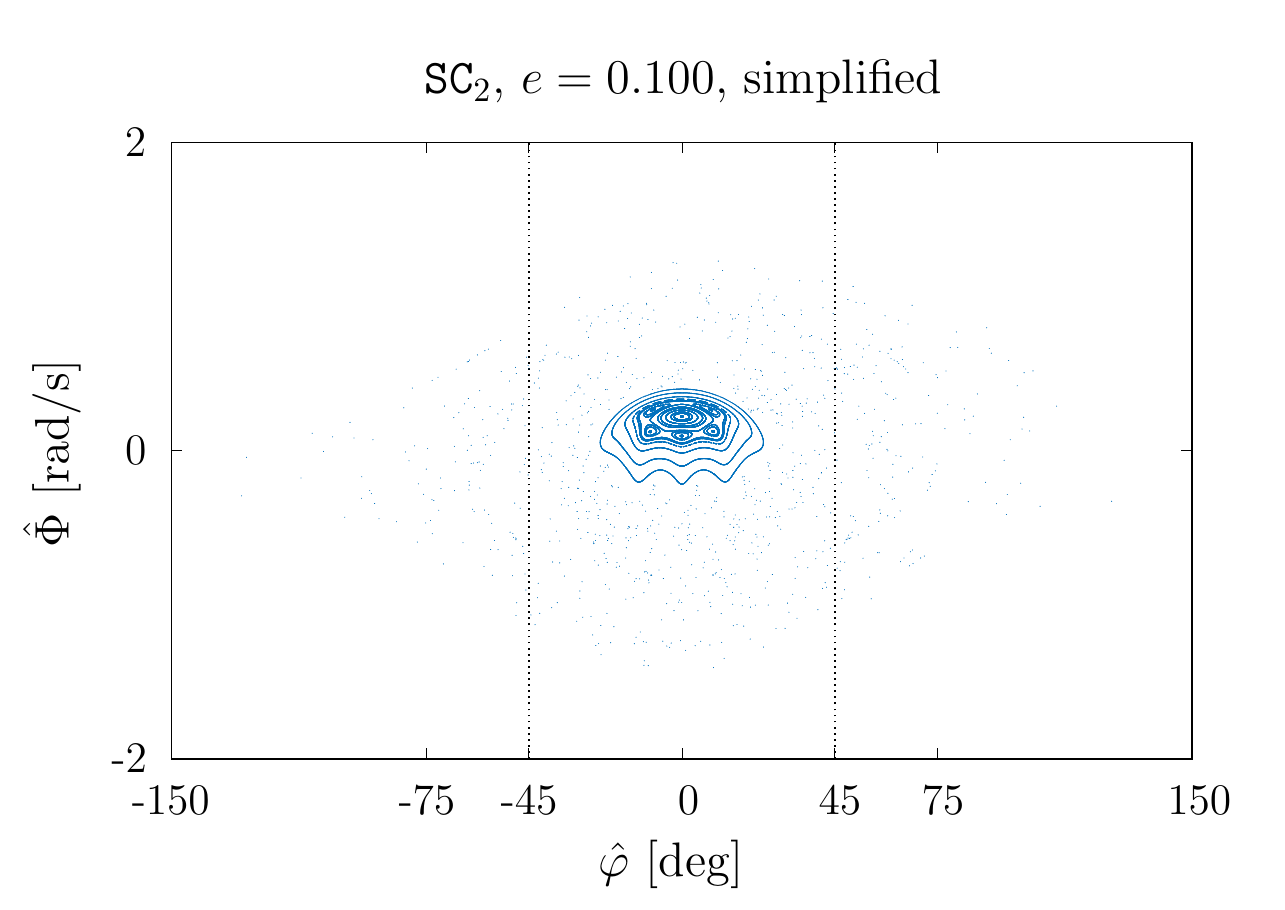}\\
\includegraphics[width = 0.48\textwidth] {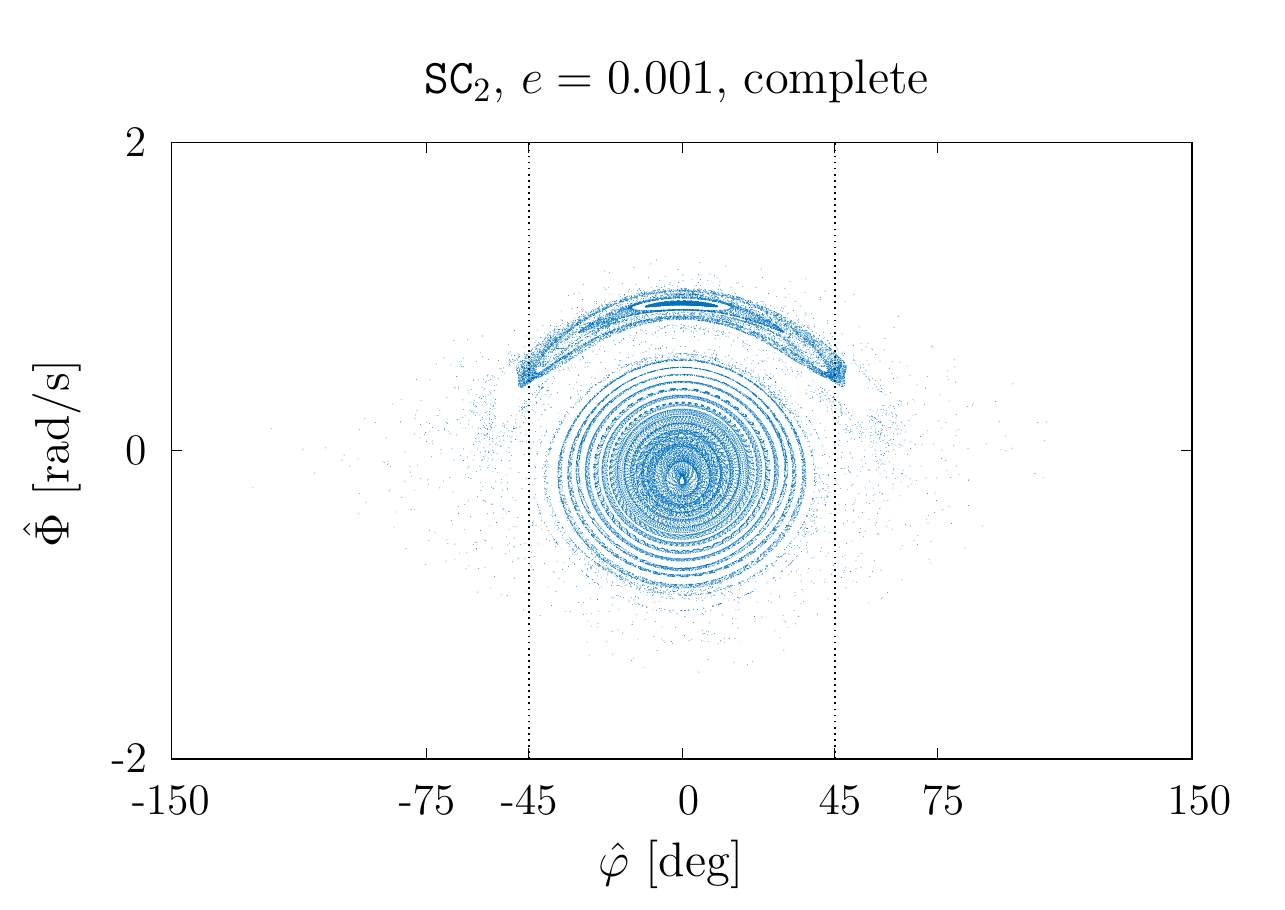}
\includegraphics[width = 0.48\textwidth] {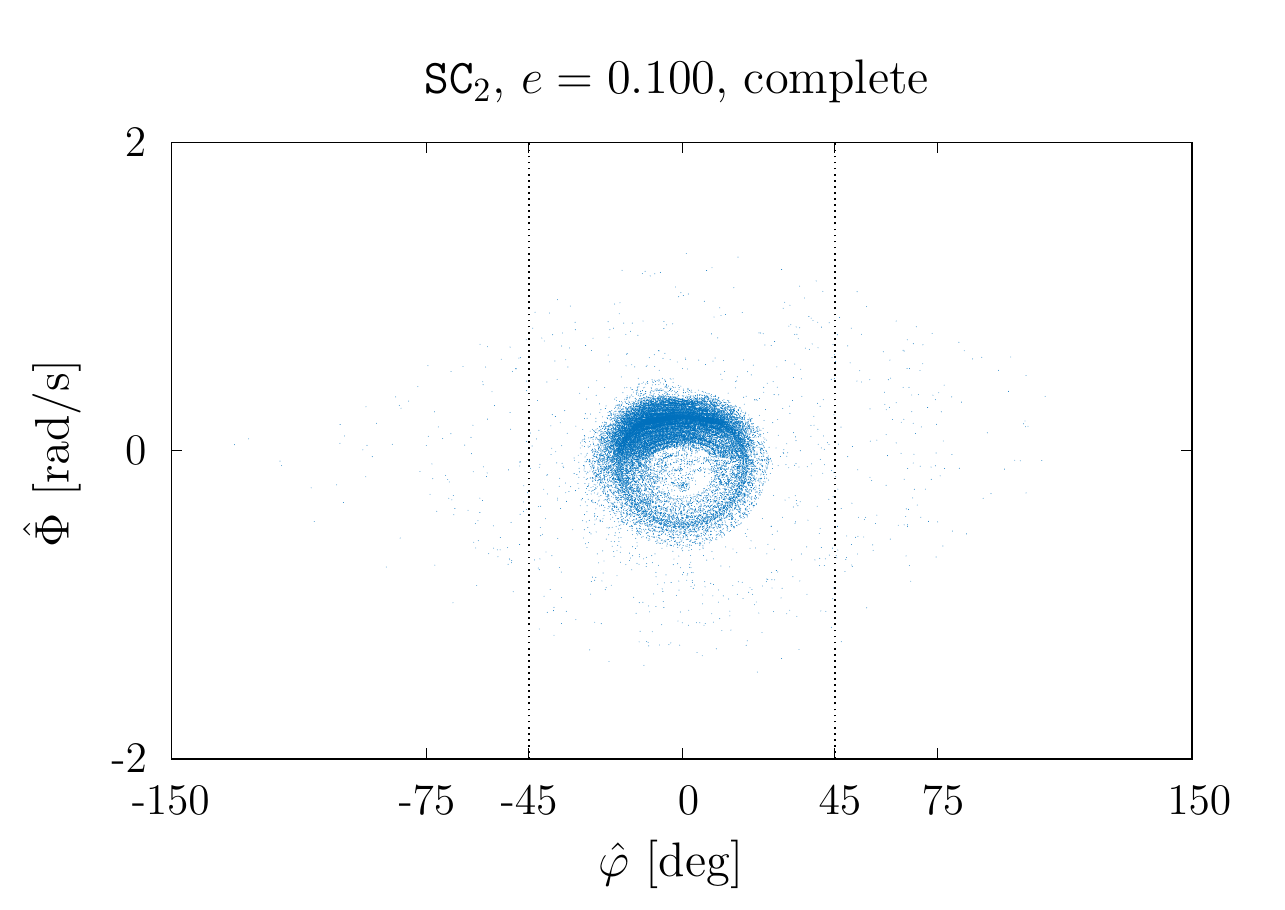}
\end{center}
\caption{Iterates of the Poincar\'e map on $\Sigma$ of the initial condition
described in \S~\ref{subsubsec:numexpSRP} for ${\tt SC}_2$. Top: simplified
model. Bottom: complete model. Left: $e=0.001$. Right: $e=0.100$.}
\label{fig:SC2-SRP}
\end{figure}

The $(\hat{\vp},\hat{\Phi})$ components of the iterates of the Poincar\'e
section to $\Sigma$ can be seen in Fig.~\ref{fig:SC1-SRP} for ${\tt SC}_1$ and
in Fig.~\ref{fig:SC2-SRP} for ${\tt SC}_2$. In both figures, the left (resp.
right) column show results for $e=0.001$ (resp. $e = 0.1$). The top (resp.
bottom) row shows results for the integration of the simplified (resp.
complete) model. In all panels isolated dots are iterates of the Poincar\'e map
that started tumbling before 1 year of integration.

% - Figura

Concerning the results for ${\tt SC}_1$ in Fig.~\ref{fig:SC1-SRP}, the is a
clear similarity between the attitude phase space of the simplified and
complete model for both values of the initial eccentricity. This due to the
fact that $d=0$ m for this spacecraft, and this is the most symmetric case
possible for the class of spacecraft taken into account. But note that there
are objects that appear in the phase space of the simplified model that are no
longer there in the complete model and vice-versa. Compare the left panels,
where orbits whose section seem to be periodic orbits close to the origin
(sun-pointing attitude) both on top and in the bottom, but in different
positions and with different periods. This can also be noticed in the right
panels. Another feature that can be seen in the right panels is that while in
the top figure there are what look like invariant curves, below it seems that
orbits fill some are in the phase space. This is more prevalent in curves that
are outside the $|\hat{\vp}|<\alpha$ region, where oscillations are wider.
There are many possible explanations of this observation. One possibility is
that this is due to the chosen section, but it can also be caused either for
the loss of differentiability of the vector field at $|\hat{\vp}|=\alpha$,
because these iterates belong to a bounded chaotic region, or even that these
initial conditions diffuse to tumbling state, but for larger time scales.

% - Figura

On the other hand, the results of the structure ${\tt SC}_2$ in
Fig.~\ref{fig:SC2-SRP} show that the displacement of the centre of mass has a
huge impact on the region of stable oscillations. Most of these are confined in
$|\hat{\vp}|<\alpha$, where the vector field is analytic. On the left we see
that the effect of the gravity gradient generated another equilibrium on top of
the central region of oscillations, and some remnant of it is still visible in
the iterates of  the complete problem. In the right panels we see that for
$e=0.100$ the region where there are stable oscillations is significatively
reduced, and in fact for the complete problem only the initial conditions
closest to the sun-pointing attitude reach 1 year without starting tumbling.

%%-----------------------------------------------------------
%% Eclipses 
%%-----------------------------------------------------------
%\subsubsection{The problem of eclipses}
%
%The fact that we are restricting ourselves to a planar problem forces makes the
%spacecraft to be in the shadow for some non-negligible time span in each
%revolution around the Earth.  In fact, for the class of spacecraft this paper
%is concerned about, only for attitude initial conditions very close to the
%sun-pointing orientation would survive these eclipses without tumbling.
%
%This is left out of the study as no dynamical solution as considering a
%moderate spin of the QRP structure (see~\cite{FHC17}) can be done when
%considering a geometry of the QRP sail when it is adapted to planar motion.
%This line of research will be tackled in forthcoming contributions devoted to
%the study of the 3D analogue of the present problem.

%---------------------------------------------------------------------- 
% Section: DRAG-dominated
%----------------------------------------------------------------------
%-----------------------------------------------------------
% Drag 
%-----------------------------------------------------------
\section{Attitude stability in a drag-dominated region}\label{sec:DRAG}

The torque and acceleration due to SRP and drag for the class of spacecraft
under consideration are similar effects in the sense that the representations
found in \S~\ref{subsubsec:SRP} and \S~\ref{subsubsec:DRAG} are the same but
the role played by the relative velocity vector in atmospheric drag is done by
sunlight direction in SRP.

Despite this similarity, the two effects are different in nature as no
reasonable simplifying assumptions - such as the apparent motion of the Sun
being perfectly circular or $p_{\rm SR}$ being constant along the motion - can
be done when studying atmospheric drag to get rid of the dependence on the
position on the orbit the spacecraft is on. Namely, atmospheric drag depends
explicitly on the orbit and on its position on it via the density $\rho$ and
the modulus of the relative velocity $v_{\rm rel}$.

\subsection{Some heuristic considerations}

A simplified deterministic model as that provided in \S~\ref{subsec:simplSRP}
can not be given in this case, yet the whole coupled attitude and orbit model
has to be tackled directly. Despite this, in light of the analysis performed in
\S~\ref{subsec:simplSRP}, some heuristic considerations can be translated in
this case to be able to draw a global description of the dynamics,
qualitatively.

On the one hand, the expected rotation dynamics of the drag sail is expected to
be oscillatory around the relative velocity vector, that now evolves as fast as
the spacecraft orbits around the Earth. For attitude initial conditions
sufficiently close to the orientation of the relative velocity vector, one
expects that if a similar exploration as that in \S~\ref{subsubsec:numexpSRP}
is performed, one could obtain an attitude phase space that is qualitatively
similar to that in Fig.~\ref{fig:phaseSRP}.  In numerical simulations, one
should plot $\hphi=\vp-\delta=\vp-\arctan(v_y/v_x)$ as abscissa and $\Phi-{\rm
d}\delta/{\rm d}t$ as ordinate. Note that we can provide an explicit expression
for the latter, as
\begin{eqnarray}\label{eq:derdelta}
\frac{{\rm d}\delta}{{\rm d}t} & = & 
\frac{1}{v_{\rm rel}^2}
\left(\frac{{\rm d}v_y}{{\rm d}t}v_x-\frac{{\rm d}v_x}{{\rm d}t}v_y\right),
\end{eqnarray}
where the derivatives of the components of the velocity are obtained by
evaluating the orbit vector field.

\begin{figure}[h!]
\begin{center}
\includegraphics[width = 0.48\textwidth]{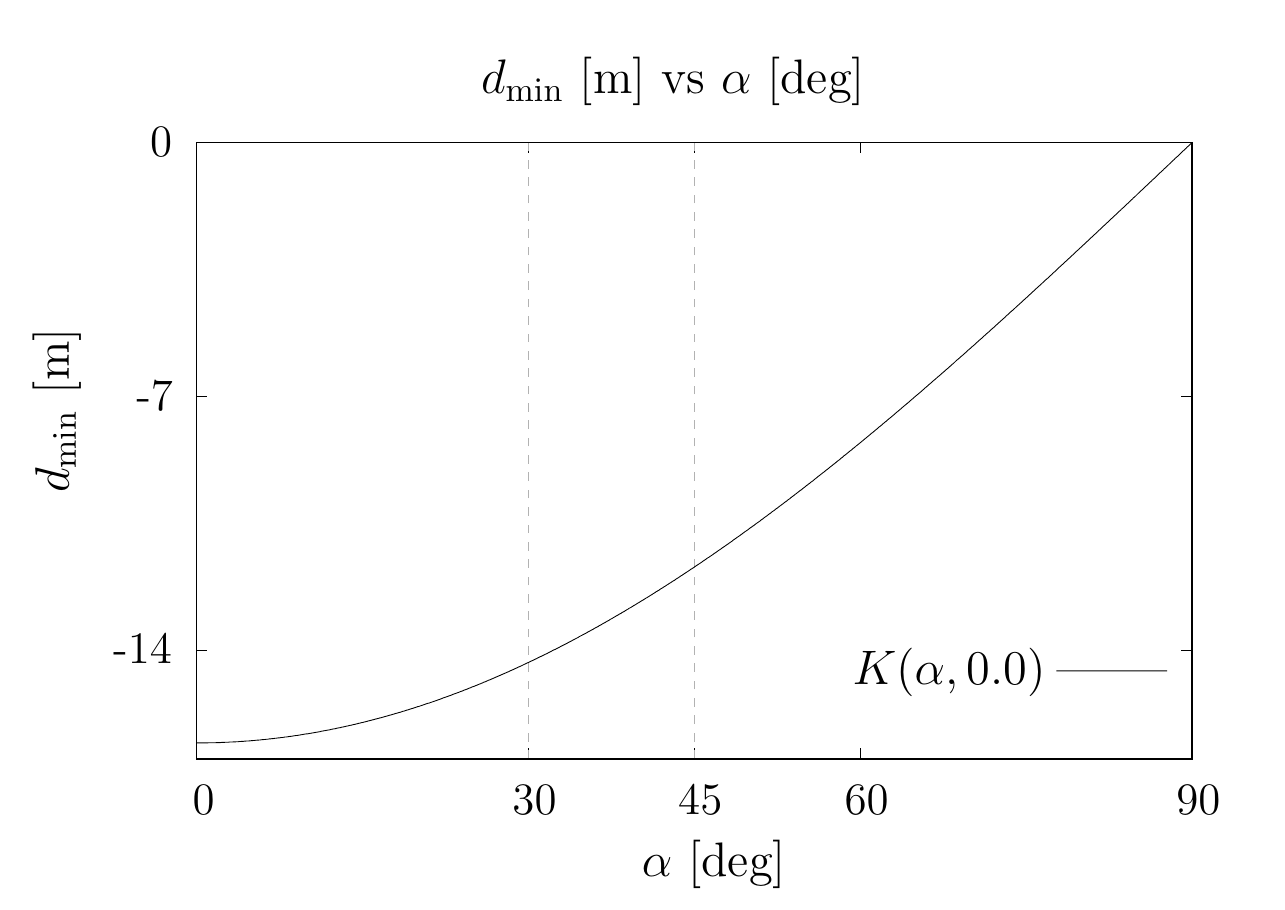}
\end{center}
\caption{Necessary conditions for the stability of the velocity-pointing
orientation of the sail in drag-dominated regions.}
\label{fig:Kdrag}
\end{figure}

On the other hand, concerning the position and stability of the attitude
equilibria, one expects the relative velocity-pointing direction to be stable,
generically, and that there is an analogous necessary stability condition
related to that for SRP that reads as Eq.~\ref{eq:stabSRP}, but setting $\eta =
0$. More concretely, the condition $k'_{1,1}=k_{1,1}(0)>0$ is equivalent to  
\begin{eqnarray}\label{eq:stabDRAG}
d > d_{\rm min} = \frac{w(m_b+m_s)}{2m_b}K(\alpha,0).
\end{eqnarray}
The right hand side of Eq.~\ref{eq:stabDRAG} is depicted in
Fig.~\ref{fig:Kdrag}, compare with Fig.~\ref{fig:dmin}. 

% - Figura

Unlike in the SRP case, where the unstable $H_\pm$ were located at $\hat{\vp} =
\pm(\pi-\alpha$); if the sail is used as a drag sail, despite analogous orbits
like $H_\pm$ exist, in the $(\vp-\delta)-(\Phi-{\rm d}\delta/{\rm d}t)$ plane, their
position depend on $\alpha$ and $d$. In particular, in case $d = 0$ m,
$k'_{0,2} = k_{0,2}(0) = 0$, see Eq.~\ref{eq:b02} and in that case
$M_0(\pm\pi/2,0)=0$, recall Eq.~\ref{eq:M0}. In this case $H_\pm$ are expected
to be close to $(\pm\pi/2,0)$. In practice, this reduces the maximal range of
oscillations that can be considered for each spacecraft, and has to be taken
into account in simulations.

%-----------------------------------------------------------
% Bounded attitude motion 
%-----------------------------------------------------------
\subsection{Bounded attitude motion}

In this subsection the performance of the sails as depicted in
Fig.~\ref{fig:shapesail} used as drag sails is tested. Unlike in the study of
these structures as solar sails in \S~\ref{sec:SRP}, where the goal was to find
attitude initial conditions that remained close to the initial state for long
periods of time, so that the average dynamics was as if the sail was flat and
always pointing to the sunlight, in this case the problem has obvious
boundaries and the performance can be measured in a well defined metric: if the
motion starts at an initial altitude $h_0$, and one considers deorbiting as
reaching a minimum value of the altitude $h_{\rm min}$, the best structure is
the one that minimizes the time of flight.

\subsubsection{Numerical experiment}\label{subsubsec:numexpDRAG}

As previously done in \S~\ref{subsubsec:numexpSRP}, since the orbit is
transversal to the axes of $\Fcal_I$, the Poincar\'e section to $\Sigma$
(recall Eq.~\ref{eq:defsigma} and the sketch in Fig.~\ref{fig:sketchSigma}) is
considered. 

To test and compare different structures in the same scenario a single orbit
initial condition is considered: in all simulations the motion starts at the
point $x=R + h_0$, $y=0$, where $R$ is the Earth's radius and $h_0 = 600$ km is
the initial altitude of the orbit. The orbit is assumed to be initially
circular, so initially $\delta_0 = \pi$ rad.

For this orbit initial condition, two illustrative numerical experiments have
been performed: First, to look for stable attitude motion, an equispaced grid
of 40 attitudes has chosen:
\begin{itemize}
	\item[-] For ${\tt SC}_1$, $\Phi_0 = {\rm d}\delta_0/{\rm d}t$ rad/s
and $\vp_0 = \delta_0 + 0.5\times 90^\circ\times j / 40$, $j = 0,\ldots, 39$, 
	\item[-] For ${\tt SC}_2$, $\Phi_0 = {\rm d}\delta_0/{\rm d}t$ rad/s
and $\vp_0 = \delta_0 + 0.5\times 135^\circ\times j / 40$, $j = 0,\ldots, 39$.
\end{itemize}
This is intended to study the performance as a function of the initial attitude
condition.\\

The second experiment consists of a study of the performance of the structure
as a function of the parameters $\alpha$ and $d$ of the structures. For each of
the values of $\alpha = 30^\circ, 40^\circ, 45^\circ, 50^\circ, 60^\circ,
70^\circ$ and $80^\circ$, two spacecraft have been considered: one with $d=0$,
and another one assuming that the bus is at the tip of the sail, that is, for
$d$ as given in Eq.~\ref{eq:busattip}. For each of these spacecraft, starting
at the same orbit initial condition as in the first experiment, we have
considered 40 attitude initial conditions, similarly chosen as
\begin{itemize}
	\item[-] For spacecraft with $d=0$ m, $\Phi = {\rm d}\delta_0/{\rm d}t$
rad/s and $\vp_0 = \delta_0 + 0.5\times 90^\circ\times j / 40$, $j = 0,\ldots,
39$,
	\item[-] For spacecraft with the bus at the tip of the sail, $\Phi =
{\rm d}\delta_0/{\rm d}t$ rad/s and $\vp_0 = \delta_0 + 0.5\times
(180^\circ-\alpha)\times j / 40$, $j = 0,\ldots,
39$, 
\end{itemize}

In all cases, ${\rm d}\delta_0/{\rm d}t$ is evaluated as in
Eq.~\ref{eq:derdelta} and $h_{\rm min} = 120$ km is considered. 

The index $j$ of the attitude initial conditions is going to be used as label
to identify them and to be able to compare results.

\subsubsection{Numerical results}

\begin{figure}[h!]
\begin{center}
\includegraphics[width = 0.48\textwidth]{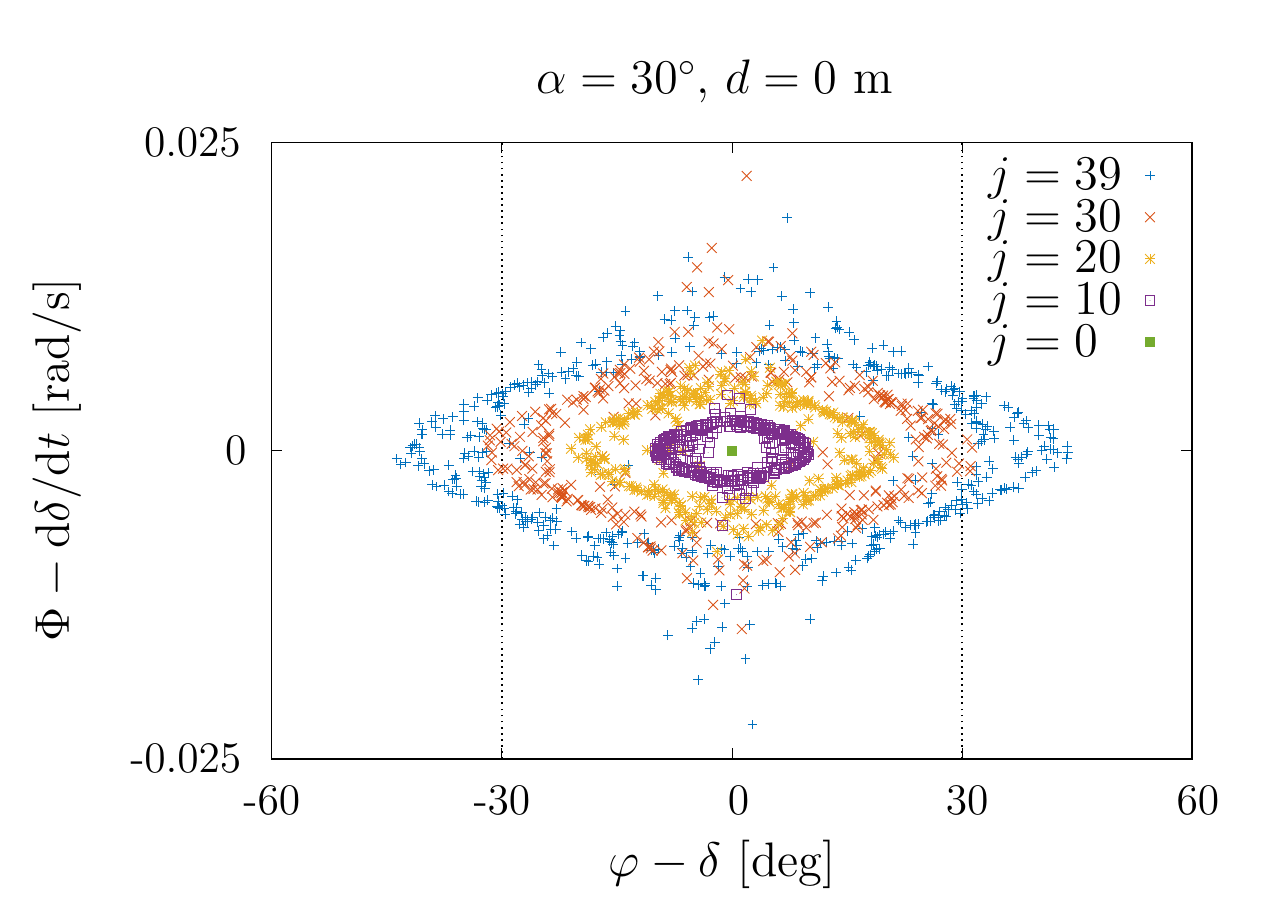}
\includegraphics[width = 0.48\textwidth]{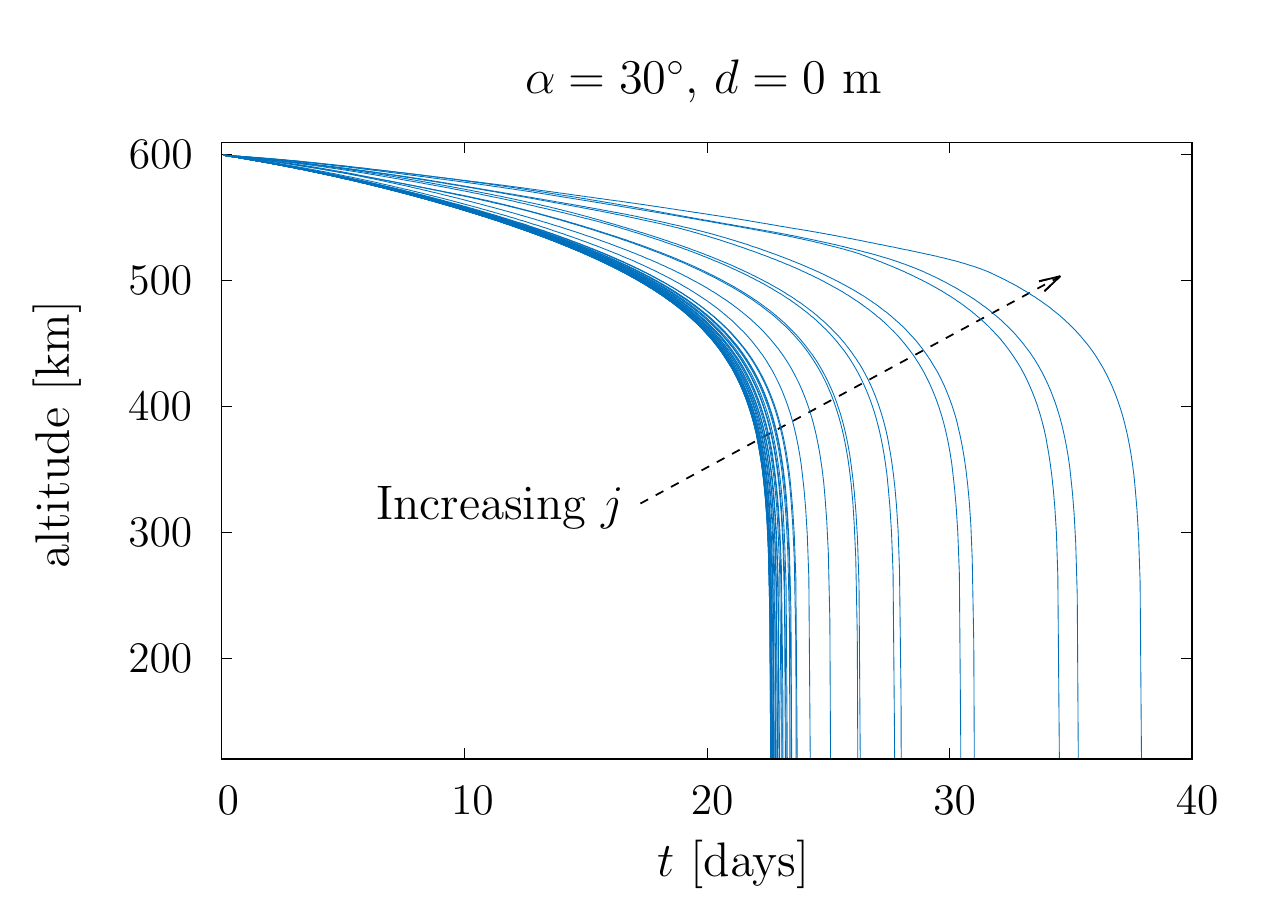}\\
\includegraphics[width = 0.48\textwidth]{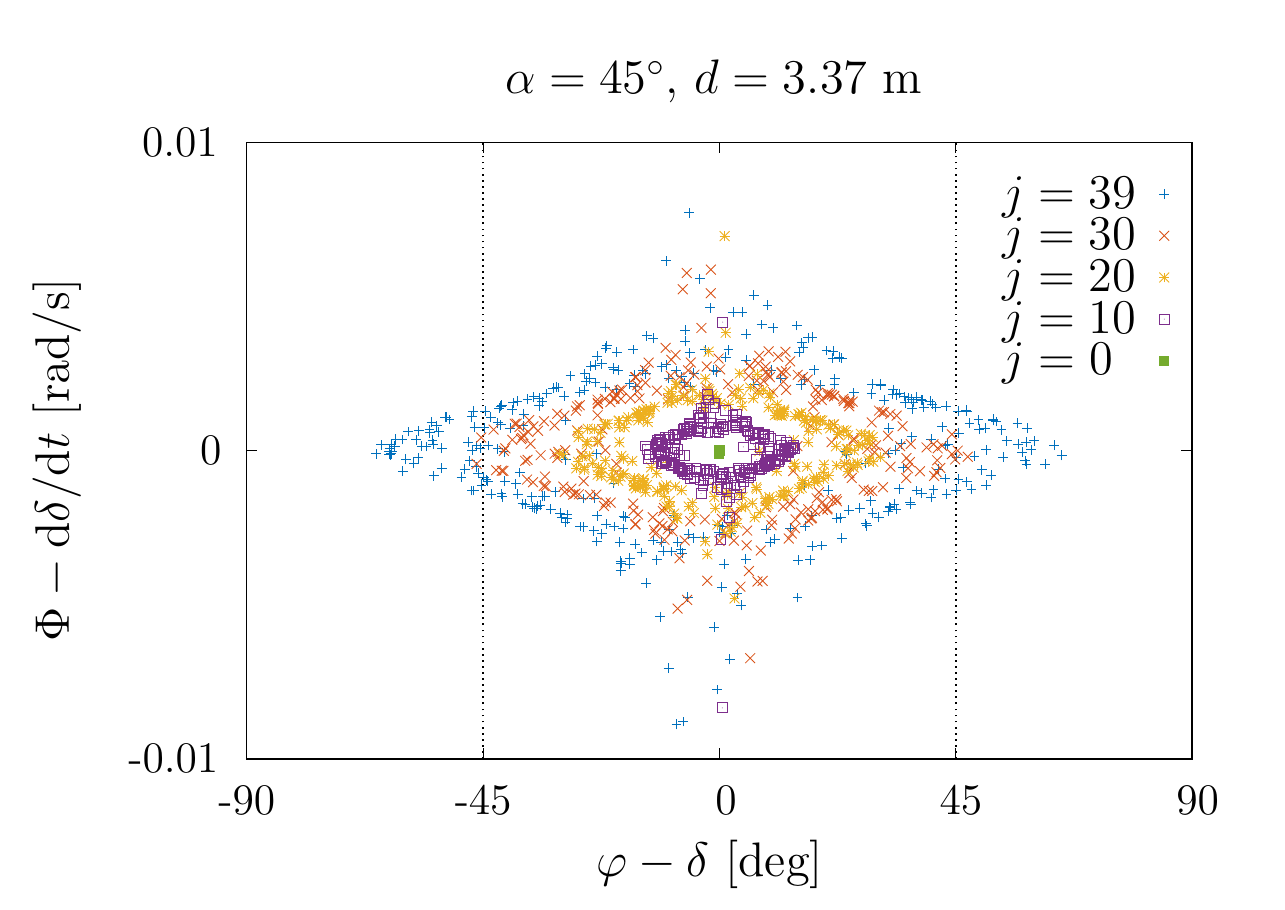}
\includegraphics[width = 0.48\textwidth]{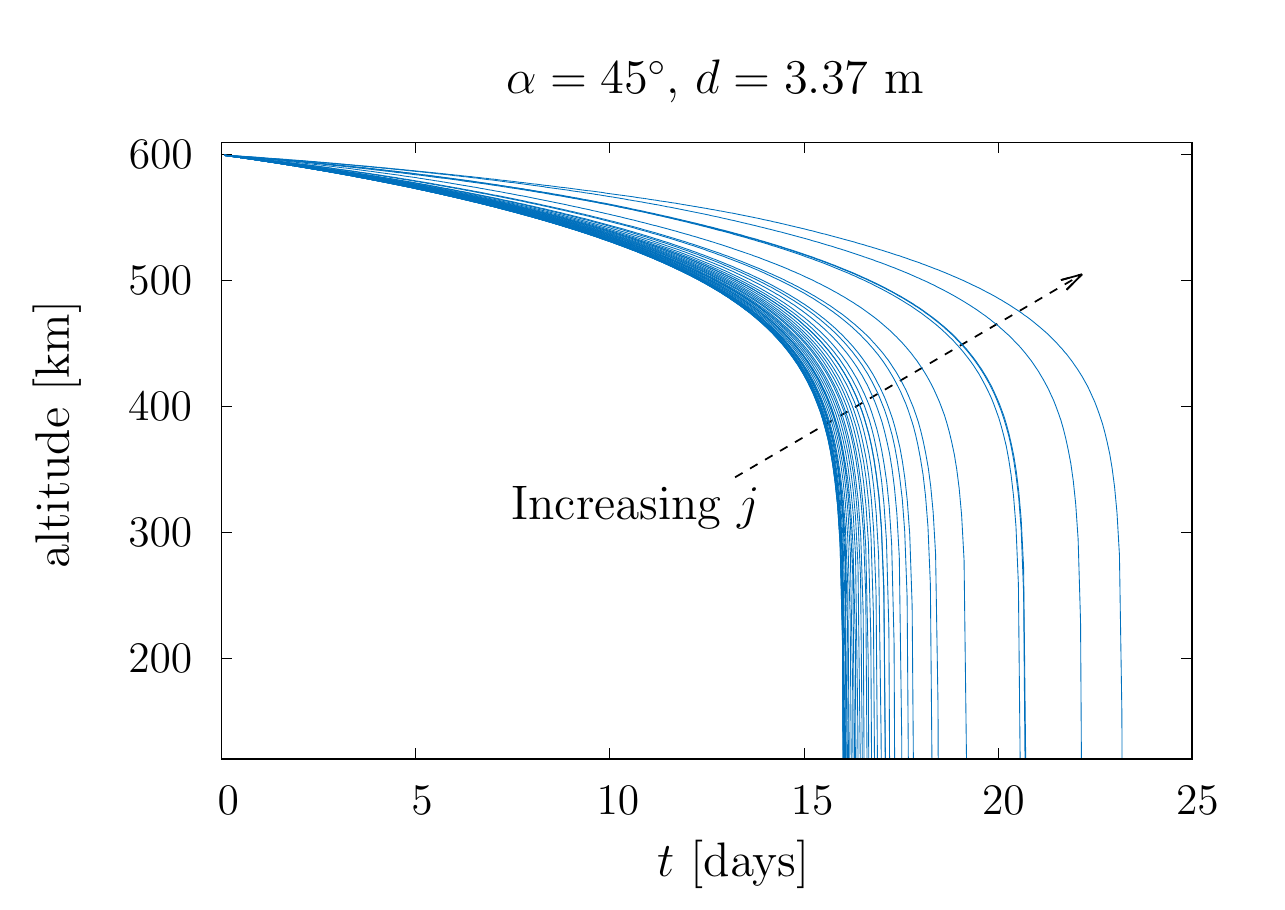}\\
\end{center}
\caption{Left: Phase space of the Poincar\'e map on the section $\Sigma$ in the
drag-dominated region. Right: Variation of the altitude of the orbit. Top:
results for ${\tt SC}_1$. Bottom: results for ${\tt SC}_2$.}
\label{fig:resDRAG}
\end{figure}

The results of the first experiment are summarized in Fig.~\ref{fig:resDRAG}.
The left panels show the attitude iterates on $\Sigma$ of five initial
conditions, those for $j=0, 10, 20, 30$ and $39$. Only five initial conditions
are displayed to be able to distinguish them easily. On the right, the
evolution of the altitude from $h_0$ to $h_{\rm min} = 120$ km for all the 40
initial conditions considered is displayed.

% - Figura

Concerning the attitude phase space in the left panels of
Fig.~\ref{fig:resDRAG}, it is remarkable how, for both spacecraft, the initial
condition $j = 0$, that starts exactly at the relative velocity-pointing
direction maintains this attitude with almost negligible variations at the
displayed scale. The rest of initial conditions seem to fill an increasing area
as time increases, in a tendency that seems to indicate slow diffusion towards
a tumbling state. In both cases, the most robust initial conditions, meaning
those whose range of $\vp-\delta$ diffuses slowlier are those where the
attitude is initially in an orientation in which both panels produce torque,
that is, where $|\vp-\delta|<\alpha$.

Concerning the deorbiting time, the displayed tendency shows that milder
oscillations lead to faster deorbiting. Comparing the right column of
Fig.~\ref{fig:resDRAG} one sees that ${\tt SC}_2$ leads to faster deorbiting
than ${\tt SC}_1$. The natural question whether this is due to the difference in
parameter $\al$ or $d$ is answered with the second experiment.\\

\begin{figure}[h!]
\begin{center}
\includegraphics[width = 0.48\textwidth]{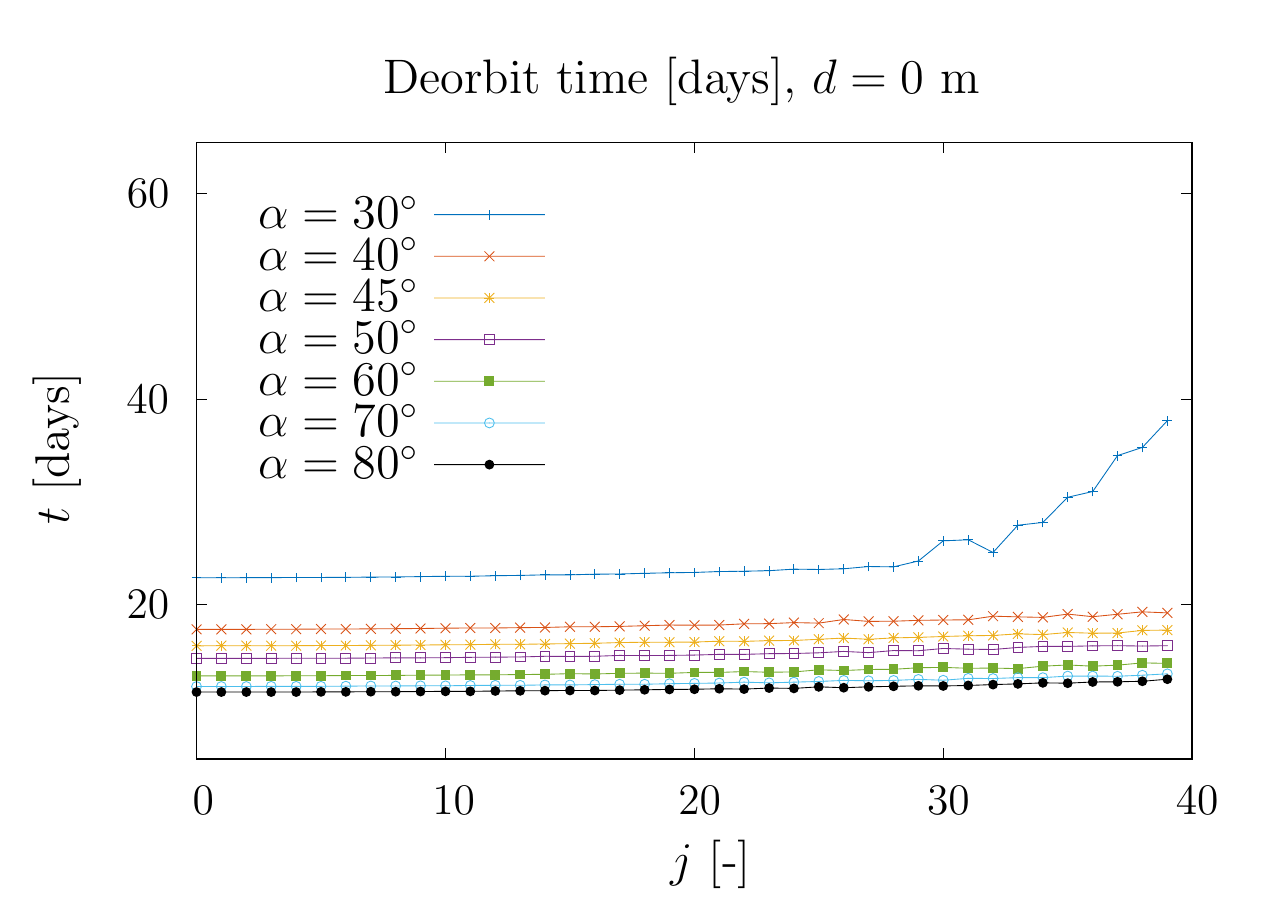}
\includegraphics[width = 0.48\textwidth]{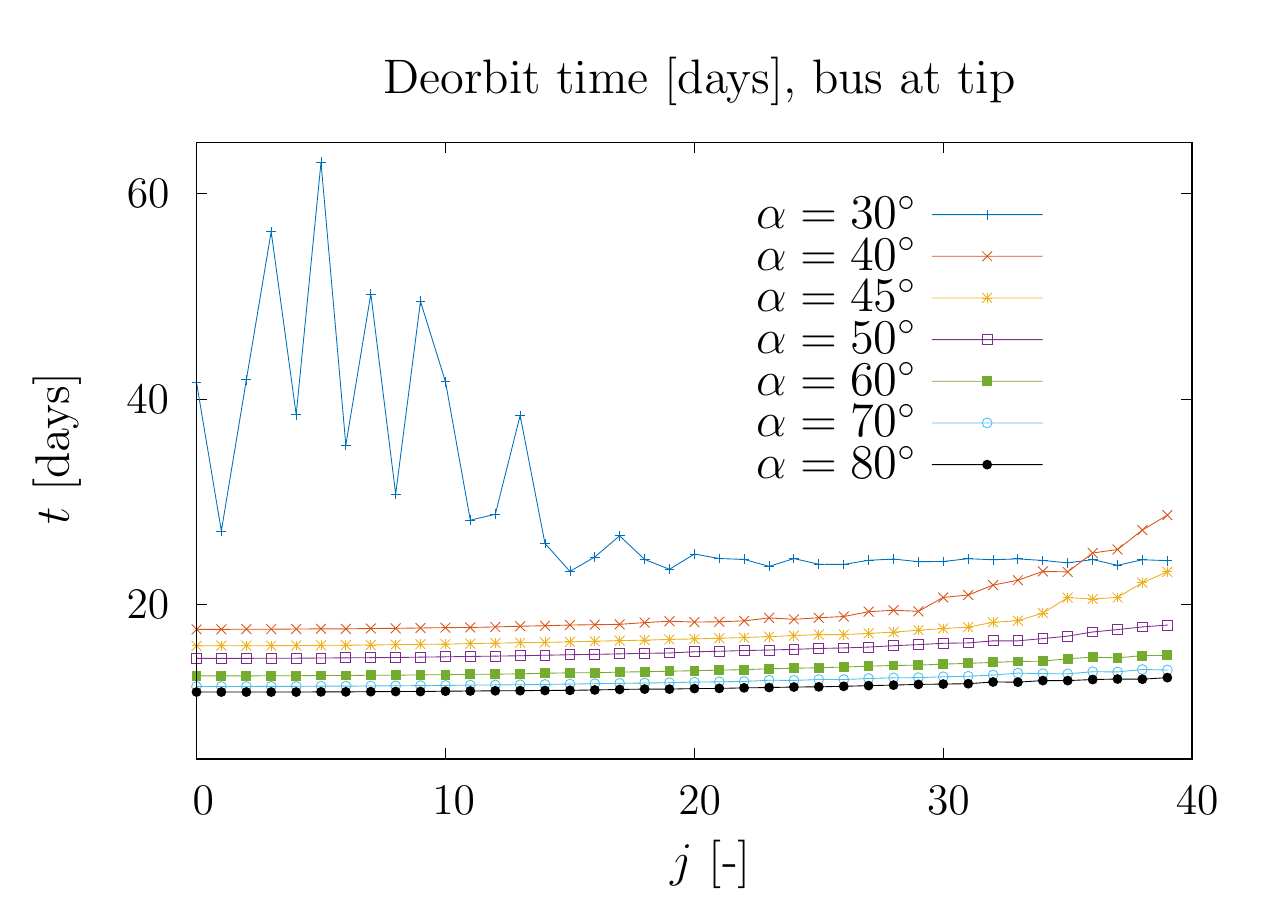}
\end{center}
\caption{Deorbit time as a function of the initial condition (distance to
velocity-pointing attitude) for different values of aperture angle $\al$. Left:
$d=0$ m. Right: $d$ chosen to be in the tip of the sail, see
Eq.~\ref{eq:busattip}.} 
\label{fig:resDRAG-tim}
\end{figure}

In Fig.~\ref{fig:resDRAG-tim} the results of the preliminary sensitivity
analysis are shown. On the left the results of spacecraft with $d=0$ m are
shown, while on the right the bus is assumed to be at the tip of the
spacecraft. In both cases one observes that, on the one hand, the larger
$\alpha$ is (meaning closer to $90^\circ$), the faster the deorbiting is. In
all cases, the larger $j$ (that is, the further one starts from the relative
velocity-pointing attitude) the longer it takes to deorbit. 

% - Figura

Also, note that smaller values of $\alpha$ give worse performance as the
amplitude of oscillations they can perform without starting tumbling is
narrower.  Finally, it is important to remark that for wider aperture angles
$\alpha$ the value of $d$ does not seem to play a leading role as the
deorbiting times are comparable.  For instance, for $\al=80^\circ$, the
deorbiting times almost coincide for small values of $j\leq 5$, with
differences below 1 min, and for larger values of $j$ in the worst case the
difference in performance for the two values of $d$ is of the order of 15 min.

%---------------------------------------------------------------------- 
% Section: Conclusions 
%----------------------------------------------------------------------
%------------------------------------------------------------
% Summary of results and conclusions 
%------------------------------------------------------------
\section{Summary and conclusions}\label{sec:conclusions}

In this work a planar reduction of the coupled orbit and attitude dynamics of a
solar/drag sail has been studied. The structures considered have a shape that
endow them with auto-stabilizing properties: the sail structure consists of
equal square panels in a way that form an angle $2\alpha$. Moreover, the
position of the payload, and hence the mass distribution of the spacecraft, is
put as a parameter, that measures the center of mass - center of pressure
offset, and this is referred to as $d$.  

One the one hand, helio-stability has been studied. The obtained results can be
summarized as follows.
\begin{enumerate}
	\item The stability of the sun-pointing attitude has been explicitly
established as an analytic expression that involves the main parameters of the
system: $\al$ and $d$. This can be used in practice as guideline for the
construction of such spacecraft. 
	\item A further simplification of the problem allows to provide a model
of the attitude motion that has Hamiltonian structure. In this model the
dynamics has been shown to be pendulum-like close to the Sun-pointing direction
and allows to measure explicitly regions of stability around this direction.
These are, in fact, candidate attitude states to oscillate around the sunlight
direction for large amounts of time in a complete problem taking into account
more effects.  
	\item The numerical results of the simplified attitude model have been
compared with numerical results of a coupled attitude and orbit system that
takes into account the $J_2$ effect and the SRP acceleration of the oscillating
sail. The better performance has been demonstrated to be for spacecraft with
smaller $d$, that is, for cases in which the gravity gradient torque is
smaller.
\end{enumerate}

On the other hand, when dealing with the sail as a drag sail, there are no
reasonable simplifying assumptions that allow to uncouple orbit and attitude
dynamics. Hence the stability analysis performed in the SRP case cannot be
directly translated into the drag case, yet some heuristic considerations can
be done in light of the previous results. Namely, a similar guideline for how
to choose $\al$ and $d$ so that the relative velocity-pointing direction is
stable can be established.

This case has the advantage that the main interest is to deorbit rather than to
maintain a concrete attitude, and the performance of the sails can be assessed
by studying deorbiting times. The results in this case can be summarized as:
\begin{enumerate}
	\item The stability of the relative velocity-pointing attitude has been
numerically demonstrated. Here in all cases the best performance is in
oscillatory motions where both panels produce torque (and hence acceleration)
through the whole motion. Moreover, the smaller the amplitude of the
oscillations, the faster deorbiting is.
	\item The dependence of the deorbiting time on the parameters $\al$ and
$d$ has been studied. The main conclusion has been that the closest $\al$ is to
$\pi/2$ rad, the better, regardless of $d$ (prescribed it is chosen so that the
velocity pointing attitude is stable).
\end{enumerate}

The presented results provide evidence of the possibilities of considering non
conventional sails such as the quasi-rhombic pyramid, as these lead to attitude
oscillating stable dynamics if treated either a solar or a drag sail. The
ranges of oscillation and the long term behaviour can be first approximated
using simplified and easily treatable models. Concerning SRP, a long-term
oscillating behaviour can be averaged out as done in~\cite{MC19}, where this
averaged dynamics was proven to be equivalent to considering a flat sail with
fixed attitude towards the Sun. When drag is the main disturbance, if the
initial state is sufficiently close to the relative velocity vector, the
deorbiting time and trajectory can be well monitored as the attitude remains
close to fixed. Similar results as those in~\cite{MC19} are expected to be
obtained in light of the results exposed in this contribution, and will appear
elsewhere.

There are a number of future lines of research that emerge from this work.  The
first and more natural one is to study the combined SRP and drag effects to
deorbit satellites~\cite{SCCM13}, but exploiting the auto-stabilizing
properties of the family of sails under consideration to reduce as much as
possible the need for attitude control. And of course the study of the dynamics
and the performance of a 3D QRP as suggested in~\cite{CHMR13, FHC17}, taking
into account the effect of eclipses.

%---------------------------------------------------------------------- 
% Agraïments 
%----------------------------------------------------------------------
%------------------------------------------------------------
% Agraiments
%------------------------------------------------------------
\section{Acknowledgements}

The research leading to these results has received funding from the Horizon
2020 Program of the European Union's Framework Programme for Research and
Innovation (H2020-PROTEC-2015) under REA grant agreement number 687500 –
ReDSHIFT. The authors also acknowledge the ERC project COMPASS (Grant agreement
No 679086), the use of the Milkyway High Performance Computing Facility and the
associated support services at the Politecnico di Milano and the use of the
computing facilities of the Dynamical Systems Group of Universiat de Barcelona.
The datasets generated for this study can be found in the repository at the
link {\tt www.compass.polimi.it/publications}. The authors also thank fruitful
conversations and support of J. Gimeno and M.  Jorba-Cusc\'o.

\bibliographystyle{plain}
\bibliography{heliodrag}
\end{document}